%% file: ICDE_main_LiFeng.tex
\documentclass[10pt,conference,letterpaper]{IEEEtran}
\usepackage{times,amsmath,epsfig}
\usepackage{graphicx}
\usepackage{subfigure}
\usepackage{algorithm}
\usepackage[noend]{algorithmic}
\usepackage{enumerate}
\usepackage[amsmath,thmmarks]{ntheorem}
\title{Challenging More Updates: Towards Anonymous Re-publication of Fully Dynamic Datasets}
\author{%
{Feng Li and Shuigeng Zhou}%
\vspace{1.6mm}\\
\fontsize{10}{10}\selectfont\itshape Department of Computer Science
and Engineering, Fudan University\\Shanghai 200433, China\\
\fontsize{9}{9}\selectfont\ttfamily\upshape \{fengli2006,
sgzhou\}@fudan.edu.cn }
\begin{document}
\maketitle

\input{0_abstract/abstract.tex}
\input{1_introduction/introduction.tex}

\input{2_dynamic/dynamic.tex}
\input{3_SUG/SUG.tex}

\input{4_principle/principle.tex}
\input{5_algorithm/algorithm.tex}
\input{6_experiments/experiments.tex}
\input{7_related/related.tex}

\input{8_conclusion/conclusion.tex}

\bibliographystyle{IEEEtran}
\bibliography{IEEEabrv,IEEEexample}
\input{9_appendix/appendix.tex}

\end{document}

%% file: 0_abstract/abstract.tex
\begin{abstract}
Most existing anonymization work has been done on static datasets,
which have no update and need only one-time publication. Recent
studies consider anonymizing dynamic datasets with \emph{external
updates}: the datasets are updated with record insertions and/or
deletions. This paper addresses a new problem: anonymous
re-publication of datasets with \emph{internal updates}, where the
attribute values of each record are dynamically updated. This is an
important and challenging problem for attribute values of records
are updating frequently in practice and existing methods are unable
to deal with such a situation.

We initiate a formal study of anonymous re-publication of dynamic
datasets with internal updates, and show the invalidation of
existing methods. We introduce theoretical definition and analysis
of dynamic datasets, and present a general privacy disclosure
framework that is applicable to all anonymous re-publication
problems. We propose a new counterfeited generalization principle
called \emph{m-Distinct} to effectively anonymize datasets with
\underline{both} external updates and internal updates. We also
develop an algorithm to generalize datasets to meet
\emph{m-Distinct}. The experiments conducted on real-world data
demonstrate the effectiveness of the proposed solution.
\end{abstract}


%% file: 1_introduction/introduction.tex
\section{Introduction}\label{section_1}

Many organizations are required to publish individual records or
other datasets for different purposes.
The released data should provide useful information as much as
possible, while the privacy issue should also be considered: any
sensitive information of individuals should not be disclosed.

For example, a hospital tends to publish a dataset of medical
records for research purpose, meanwhile it does not hope to reveal
any sensitive information to public. Table~\ref{microdata_T1}
illustrates such an original dataset.
Apparently, the ``Name'' attribute, which explicitly indicates an
individual (called \emph{identifier}), should be hidden from the
public. Moreover, the other non-sensitive attributes (``Zipcode'',
``Hours/week'') should not be published directly either. Because
with the help of background knowledge, an adversary may identify an
individual by the combination of these attribute values. This kind
of attacks and the combination of these attributes are often
referred to as \emph{linking attack} and
\emph{quasi-identifiers}(QI) respectively.

\emph{Generalization}~\cite{samarati:anonymity} is a prevailing
technique that can be exploited to anonymize datasets and protect
sensitive information. It hides the specific attribute values by
publishing less specific forms of QI attribute values. Since several
individual records may have the same generalized attribute values,
which will causes these records indistinguishable. We call the
published records that have the same generalized QI attribute value
a \textit{QI-group}. Besides, generalization is presence
resistant~\cite{presence} in a certain degree as it does not publish
the accurate QI attribute values directly. This feature makes itself
superior to \textit{anatomy}~\cite{anatomy} et al. in some extent.

\subsection{Motivation}
\makeatletter
\newcommand\figcaption{\def\@captype{figure}\caption}
\newcommand\tabcaption{\def\@captype{table}\caption}
\makeatother

\begin{figure*}[htb]
\begin{minipage}[b]{0.28\textwidth}
\small{ \centering\tabcaption{Microdata
$T_1$}\label{microdata_T1}\vspace{-1em}
\begin{tabular}{c|c|c|c|} \cline{2-4}

\bf{Name} & \bf{Zip.} & \bf{H} &\bf{Disease}\\
\hline

Ken & 14k & 20& Dyspepsia\\\cline{2-4}

Julia & 16k & 23& Pneumonia\\\cline{2-4}

Tom & 24k & 32& Pneumonia\\\cline{2-4}

Harry & 26k & 35& Gastritis\\\cline{2-4}

Lily & 29k & 17& Glaucoma\\\cline{2-4}

Ben & 31k & 19& Flu\\\cline{2-4}
\end{tabular}
\vspace{-1em} }
\end{minipage}
\begin{minipage}[b]{0.29\textwidth}
\small{ \centering \tabcaption{Microdata
$T_2$}\label{microdata_T2}\vspace{-1em}\begin{tabular}{c|c|c|c|}
\cline{2-4} \bf{Name} & \bf{Zip.} & \bf{H} &\bf{Disease}\\ \hline

Ken & 14k & 20& Dyspepsia\\\cline{2-4}

Julia & \underline{\emph{18k}} & \underline{\emph{31}}&
\underline{\emph{Lung Cancer}}\\\cline{2-4}

Tom & \underline{\emph{15k}} & \underline{\emph{27}}&
Pneumonia\\\cline{2-4}

Harry & \underline{\emph{23k}} & \underline{\emph{32}}&
\underline{\emph{Dyspepsia}}\\\cline{2-4}

Lily & \underline{\emph{12k}} & 17& Glaucoma\\\cline{2-4}

Ben & \underline{\emph{26k}} & \underline{\emph{35}}&
\underline{\emph{Pneumonia}}\\\cline{2-4}

\end{tabular}
\vspace{-1em} }
\end{minipage}
\begin{minipage}[b]{0.4\textwidth}
\small{ \centering \tabcaption{Generalization
$T^*_1$}\label{Generalization_T1}\vspace{-1em}\begin{tabular}{c|c|c|c|c|}
\cline{2-5} \bf{Name} &\bf{GID}& \bf{Zip.} & \bf{H} &\bf{Disease}\\
\hline

Ken & 1&[14k, 16k] & [20, 23]& Dyspepsia\\\cline{2-5}

Julia& 1& [14k, 16k] & [20, 23]& Pneumonia\\\cline{2-5}

Tom & 2& [24k, 26k] & [32, 35]& Pneumonia\\\cline{2-5}

Harry & 2&[24k, 26k] & [32, 35]& Gastritis\\\cline{2-5}

Lily&  3&[29k, 31k] & [17, 19]& Glaucoma\\\cline{2-5}

Ben &  3&[29k, 31k] & [17, 19]& Flu\\\cline{2-5}

\end{tabular}
\vspace{-1em} }
\end{minipage}
\end{figure*}

Most existing anonymization researches have focused on static
datasets. However, real datasets are dynamic. These datasets are
usually updated frequently, thus re-publication is required.
Anonymizing and re-publishing dynamic datasets is a challenging
task. Not only is the increasing number of publication times
required, but also both of the old and new sensitive information
need to be well protected.

The complexity of dynamic datasets anonymization is caused by data
updates. We can classify dynamic dataset updates to two types:
\emph{external update} and \emph{internal update}. Intuitively,
external update is the update of the records in a dataset, e.g.,
record insertion and deletion will cause external update as the
total records in the dataset are not the same as before.
%and was first studied by Xiao and
%Tao~\cite{xiao:m-invariance}.
Internal update is the update of each
record's attribute values. In other words, in a dynamic dataset with
internal updates, the attribute values of each record may be
dynamically updated. For example, as a person's age grows, her/his
salary may increase.
In addition, we have the following observation about internal
updates:

\begin{figure*}[htb]
\begin{minipage}[u]{0.42\textwidth}
\small{ \centering \tabcaption{Generalization
$T^*_2$}\label{Generalization_T2}\vspace{-1em}\begin{tabular}{c|c|c|c|c|}
\cline{2-5} \bf{Name} &\bf{GID}& \bf{Zip.} & \bf{H} &\bf{Disease}\\
\hline

Ken & 1&[12k, 14k] & [17, 20]& Dyspepsia\\\cline{2-5}

Lily &1& [12k, 14k] & [17, 20]& Glaucoma\\\cline{2-5}

Julia & 2&[15k, 18k] & [27, 31]& \emph{Lung Cancer}\\\cline{2-5}

Tom &2&[15k, 18k] & [27, 31]&Pneumonia\\\cline{2-5}

Harry & 3&[23k, 26k] & [32, 35]& \emph{Dyspepsia}\\\cline{2-5}

Ben &3&[23k, 26k] & [32, 35]& \emph{Pneumonia}\\\cline{2-5}

\end{tabular}
\vspace{-2em} }
\end{minipage}
\begin{minipage}[u]{0.42\textwidth}
\centering \tabcaption{Counterfeit Generalization
$T^*_2$}\label{CG_T2}\vspace{-1em}
\scriptsize{\begin{tabular}{c|c|c|c|c|} \cline{2-5} \bf{Name} &
\bf{GID}& \bf{Zip.} & \bf{H} &\bf{Disease}\\ \hline Ken & 1&[14k,
15k] & [19, 27]& Dyspepsia\\\cline{2-5}

Tom & 1&[14k, 15k] & [19, 27]& Pneumonia\\\cline{2-5}

Julia & 2&[18k, 23k] & [31, 32]& \emph{Lung Cancer}\\\cline{2-5}

Harry &2&[18k, 23k] & [31, 32]& \emph{Dyspepsia}\\\cline{2-5}

Lily & 3&[10k, 12k] & [16, 17]& Glaucoma\\\cline{2-5}

\emph{$c_1$} & 3&[10k, 12k] & [16, 17]&
\emph{Pneumonia}\\\cline{2-5}

Ben & 4&[26k, 27k] & [35, 37]& \emph{Pneumonia}\\\cline{2-5}

\emph{$c_2$} & 4& [26k, 27k] & [35, 37]&
\emph{Cataract}\\\cline{2-5}
\end{tabular}
\vspace{-2em}

}
\end{minipage}
\begin{minipage}[u]{0.12\textwidth}
\small{\centering \tabcaption{Counterfeit
Statistics}\label{CG_T2p}\vspace{-1em}
\begin{tabular}{|c|c|} \hline
\bf{GID} & \bf{Count}\\ \hline

3 & 1\\\hline 4&1\\\hline
\end{tabular}
\vspace{-2em} }
\end{minipage}
\end{figure*}

\newtheorem{Observation}{Observation}
\begin{Observation} In a dataset, the updates of
attribute values are seldom arbitrary: there are certain
correlations between the old value and the new one.
\end{Observation}

For example, a person's current highest degree is ``bachelor'';
several years later, although we can not determine her/his highest
degree without complementary knowledge, we can conclude that it will
not be lower than ``bachelor'' and will be one of \{``Bachelor'',
``Master'', ``PHD.''\} with different non-zero probabilities.

Based on the observation above, in this paper we assume that all
updates on sensitive values are not arbitrary\footnote{As explained
later, if all updates on sensitive attribute values are random, this
dataset can be treated as a static one in anonymization.}. The
possible updates and their probabilities are estimable, and can be
treated as background knowledge known to public.

To demonstrate the challenges brought by internal updates, we give
an example with only internal updates as follows. Note that the
challenges remain when external and internal updates coexist.

Consider a hospital that carries out a project of tracking disease
evolution. Every two months, it releases medical records of the same
group of patients to other institutes; meanwhile, it also hopes to
preserve the patients' privacy.

The original microdata of the $1^{st}$ and $2^{nd}$ releases are
shown in Table~\ref{microdata_T1} and Table~\ref{microdata_T2}. In
each table, there are 6 records and each one corresponds to a unique
patient. In the $2^{nd}$ release, some attribute values of the
records (underlined) are updated.
\subsubsection{Invalidation of l-diversity}\label{Invalid-l}
We take \textit{l}-diversity to illustrate the invalidation of
existing publication solutions, and the others are similar. Briefly,
\textit{l}-diversity requires that every QI-group should contain at
least \textit{l} ``well-presented'' sensitive values. One simple
interpretation is ``distinct'', which means there are at least
\textit{l} distinct sensitive values in each QI-group.
%For instance,
%table \ref{Generalization_T1} is \textit{2}-diverse, because there
%are at least 2 distinct sensitive values in every QI-group.

Table~\ref{Generalization_T1} and Table~\ref{Generalization_T2} are
the published data of the $1^{st}$ and $2^{nd}$ releases
respectively\footnote{Actually, published tables do not contain the
identifier attribute ``Name'', we keep it here just for the
convenience of explanation.}, both are \textit{2}-diverse.
\textit{2}-diversity ensures that an adversary can not determine the
exact disease of each patient if ignoring the correlation between
the two releases. However, in practice, the situation can get worse.

For example, suppose an adversary knows that the medical records of
Ben are in both releases.  Furthermore, s/he also knows his detail
information of each time\footnote{The information can be acquired
from many sources such as voter list.}: $<$\textit{Ben, 31k, 19}$>$
in the $1^{st}$ release and updated to be $<$\textit{Ben, 26k,
35}$>$ in the $2^{nd}$ release. The adversary will reason as
follows: Ben must be in group 3 of both releases. The diseases he
may contract are in \{\textit{Glaucoma, Flu}\} and
\{\textit{Dyspepsia, Pneumonia}\} respectively. The adversary knows
that, although Ben's disease may be different in the two releases,
there must be correlation between them. Since glaucoma can not
update to both dyspepsia and pneumonia, the adversary concludes that
Ben must contract flu in the $1^{st}$ release; both of glaucoma and
flu can not update to dyspepsia, s/he can determine that Ben
contracts pneumonia in the $2^{nd}$ release.

By exploiting the correlation between the two releases, the
adversary can also disclose more sensitive information such as the
disease of Ken and Lily in both releases, the disease of Julia in
the $1^{st}$ release etc.

\subsubsection{Invalidation of m-Invariance}
\textit{m}-Invariance~\cite{xiao:m-invariance} was proposed to
re-publish dynamic dataset with only external updates. It achieves
anonymization by ensuring that in each release, the QI-group to
which an arbitrary record belongs always has the same set of
sensitive values. However, if there are internal updates in the
dataset, the requirement of \textit{m}-Invariance may be never met.

Suppose in the $1^{st}$ release Julia is in a QI-group of which the
set of sensitive values is \{\textit{Dyspepsia, Pneumonia}\}. Later,
the disease she contracted is deteriorated into lung cancer, which
will lead to that in the $2^{nd}$ release. The QI-group Julia is in
can never contain the same set of sensitive values, for the QI-group
must contain lung caner, which is not covered by
\{\textit{Dyspepsia, Pneumonia}\}, thus the requirement of
\textit{m}-Invariance is unreachable.

\subsection{Contributions}\label{motiv-distinct}
The internal updates causes the ineffectiveness of existing
solutions in privacy preservation, because internal updates can
enhance the adversary's background knowledge and shrink the scope of
an individual's possible sensitive values. The situation will get
worse as more publications are released, which will provide an
increasing amount of background knowledge.

Let us revisit the previous example. By using our solution in this
paper, Table~\ref{CG_T2} and \ref{CG_T2p} will be published, instead
of Table~\ref{Generalization_T2} for the $2^{nd}$ release. Eight
records in Table~\ref{CG_T2} (including 2 counterfeit records $c_1$
and $c_2$) are partitioned into 4 QI-groups. Table~\ref{CG_T2p}
contains the counterfeit statistics of Table~\ref{CG_T2}.

Reconsider that an adversary attempts to disclose the disease of
Ben. S/he knows that Ben must in group 3 and group 4 of the two
releases respectively. However, s/he can not determine the exact
diseases Ben contracted in both releases. Because glaucoma may
update to be cataract, and flu may update to be pneumonia: the two
possible diseases in the $1^{st}$ release can not be excluded even
exploiting the correlation between the two releases. Similarly, s/he
can not exclude any possible disease of Ben in the $2^{nd}$ release
either. Although Table~\ref{CG_T2p} indicates that there are
counterfeit records in the $2^{nd}$ release, it provides no help for
the adversary to exclude any possible disease of Ben.

The core idea of our solution is to maintain the
indistinguishability of the sensitive values in each QI-group
persistently, even though there are internal updates and the
adversary exploits the correlation between different releases.
In each release, we partition each individual's record into a
QI-group that will not lead to any exclusion of its possible
sensitive values. We also exploit counterfeit records if not enough
records to form such a QI-group for an individual.

In this paper, we initiate a formal study on the anonymization of
dynamic datasets with \underline{both} internal and external
updates. Internal updates lead to quite different challenges to
anonymization of dynamic datasets from that of external updates, and
invalidate all existing solutions. To the best of our knowledge,
this is the first work to study internal updates problem.

We first give a formal description of dynamic datasets and updates
(Section~\ref{section_2}). We then propose a novel privacy
disclosure framework called SUG (Section~\ref{section_3}), which is
applicable to all anonymous re-publication problems. By exploiting
SUG, we show how the inference works and how to estimate the
disclosure risk of sensitive information. Following that, we
introduce a counterfeit generalization principle called
\textit{m}-Distinct (Section~\ref{section_4}) to securely anonymize
and re-publish dynamic datasets with both internal and external
updates. An algorithm is also developed to achieve
\textit{m}-Distinct generalization (Section~\ref{section_5}).
Finally, experiments are conducted on real-world data to show the
inadequacy of existing solutions and the effectiveness of our
solution (Section~\ref{section_6}).

%% file: 2_dynamic/dynamic.tex
\section{Theoretical Foundation}\label{section_2}
Consider that $T$ is the microdata table that needs to be published,
it has an identifier attribute $ID$, $m$ QI attributes
$Q=\{Q_1,Q_2,...,Q_m\}$ and a sensitive attribute\footnote{In this
paper, we focus mainly on discrete sensitive attributes, since
continuous values can be discretized by various methods.} $S$. Each
record $t$ is organized as $<id, q_1, q_2,..., q_m, s>$. For an
attribute $\mathcal {A}$, record $t$'s value on $\mathcal {A}$ is
represented as $t[\mathcal {A}]$. We denote the generalized table by
$T^*$ and the generalized record by $t^*$.
If several records in $T^*$ have the same generalized QI values,
these records form a QI-group $g$. If record $t$ is in QI-group $g$,
we denote $t$'s \textit{candidate sensitive set} $C$ as the set of
sensitive values in $g$.

Let $i$ be the timestamp, and $T_i$ and $T^*_i$ be the microdata
table and generalized table of the $i^{th}$ publication,
respectively. If in $T_i$, there is a record $t_i$ ($t_i \in T_i$)
such that $t[ID]=t_i[ID]$, we say $t_i$ is $t$'s $i^{th}$ version.
Generally, we say two records which appears in different
publications are the \emph{different} versions of the \emph{same}
record, if they have the same $ID$ attribute value.

%If a record $t$ has not been deleted from $T_i$, we say
%$t_i$~($t[ID]=t_i[ID]$) is its $i^{th}$ version and $t_i \in T_i$.
%In the following section, we say two records are the \emph{same}
%record if they have similar $ID$ attribute value.

\subsection{Dynamic Dataset}

Generally, a dataset is dynamic iff its data is different at
different time. The differences are due to two types of updates:
\emph{external update} and \emph{internal update}.

\newtheorem{definition}{Definition}
\begin{definition}[External Update]
For any integer $i$ and $j$~($0\leq i<j$), if a record $t$~($t\neq
\phi$) satisfies one of the following conditions:
\begin{enumerate}
\item $t_i\in T_i$ and $t_j\notin T_j$.\label{condition_delete}
\item $t_i\notin T_i$ and $t_j\in T_j$.\label{condition_insert}
\end{enumerate}
We say that $t$ is an external update of $T_j$ in contrast to $T_i$.
\end{definition}

Based on the definition above, there are two types of specific
external updates: \emph{insertion} and \emph{deletion}, which
correspond to condition~\ref{condition_insert}
and~\ref{condition_delete}, respectively.
Another type of update, which occurs inside records, is called
\textit{internal update}.

\begin{definition}[Internal Update]
For integer $i$ and $j$~($0\leq i<j$), suppose $t_i\in T_i$ and
$t_j\in T_j$ holds for a record $t$. If $t_i$ and $t_j$ satisfy at
least one of the following conditions:
\begin{enumerate}
\item $t_i[Q] \neq t_j[Q]$.
\item $t_i[S] \neq t_j[S]$.
\end{enumerate}
Then we say that there are internal updates on $t$ in the period of
[$i$, $j$].
\end{definition}

Internal updates may occur on either QI attribute values or
sensitive attribute values. Especially, updates on sensitive
attribute values will bring more difficulty to the anonymization of
dynamic datasets as they enhance the adversary's background
knowledge about sensitive information.

\newtheorem{example}{Example}
Following external update definition, we have the definition of
external dynamic datasets as follows.

\begin{definition}[External Dynamic Dataset]
For integer $i$ and $j$~($0\leq i<j$), if dataset $T$ has the
following properties:
\begin{enumerate}
\item $\exists$ $i$, $j$, $T_j$ is externally updated in contrast to $T_i$.
\item $\forall$ $i$, $j$, if $t_i\in T_i$ and $t_j\in T_j$ holds for any record $t$, then $\forall$ $\mathcal{A}$, $t_i[\mathcal {A}]=t_j[\mathcal
{A}]$ must holds.
\end{enumerate}
Strictly, we say that $T$ is external dynamic.
\end{definition}

Intuitively, if there are record insertions and/or deletions, and
the attribute values of each record will not change as time goes,
the dataset is external dynamic. Similarly, we have the formulation
of \textit{internal dynamic dataset} as follows.

\begin{definition}[Internal Dynamic Dataset] For integer $i$ and $j$~($0\leq i<j$),
if $T$ has the following properties:
\begin{enumerate}
\item $\exists$ $i$, $j$, and internal updates that happen on a record in the period of [$i$, $j$].
\item $\forall$ $i$, $j$, $T_i$ and $T_j$ has the \emph{same} records.
\end{enumerate}
Strictly, we say that $T$ is internal dynamic.
\end{definition}

Anonymization of internal dynamic datasets has never been addressed
in the literature. In this paper, we deal with \textit{fully dynamic
datasets}, which contain both external and internal updates.

\begin{definition}[Fully Dynamic Dataset] For integer $i$ and $j$~($0\leq i<j$),
if dataset $T$ has at least one of the following properties:
\begin{enumerate}
\item $\exists$ $i$, $j$, and $T_j$ that has external updates in contrast to $T_i$;
\item $\exists$ $i$, $j$, and internal update(s) occurring on at least one record during [$i$, $j$].
\end{enumerate}
Then, we say that the dataset is fully dynamic.
\end{definition}

Both of the internal dynamic dataset and external dynamic dataset
are special cases of fully dynamic dataset as fully dynamic dataset
may be updated by internal updates and external updates. As
explained in~\cite{xiao:m-invariance}, external update brings
\textit{critical absence} and other challenges into dynamic dataset
anonymization. However, Internal update will leads to different
challenges comparing to the previous work.

First, there will be an $i^{th}$ version $t_i$ for an individual's
record $t$ if it has not been removed from $T_i$. Furthermore, each
version is an access to the record's sensitive information and the
total amount is increasing as time evolves. This is different from
the situations in static dataset and external dynamic dataset, which
always have immobilizing record for an individual.

Second, there is correlation between different versions of an
individual's record. In other words, the series of internal updates
on an individual's record are not independent. That makes the
situation more complex: in contrast to external update, the record
insertion or deletion are usually independent\footnote{Actually,
this is an implicit assumption the previous work~
\cite{xiao:m-invariance,byun:incremental} makes.}.

Third, the sensitive attribute value will be updated by internal
update, which means the current one may be different from the
historical values. Thus when publishing a dataset, both of the
historical sensitive values and the current one need be well
protected. The situation will get worse as the increasing releases
of the individual's sensitive value.

Forth,one breach of sensitive value may lead to chain-action breach
if exploiting their correlation~\cite{corruption}.

\subsection{Problem Formulation}
Additional background knowledge rose by the updates promotes the
disclosure probability of sensitive information. In this paper we
classify the background knowledge into two types: \textit{explicit
background knowledge} is specifically related to the publication of
the dataset while \textit{implicit background knowledge} is more
general.

\begin{definition}[Explicit Background
Knowledge]\label{def_explicit_bk}$\\$ i) For any positive integer
$i$, there is an \emph{external knowledge table}
$E_i=<ID,Q_1,Q_2,...,Q_m>$ corresponding to $T_i$, which contains
the ID attribute and QI attributes data of $T_i$. $\\$ii) In the
span of $[1, n]$, we denote the union of the published tables as
$PT_n=\bigcup_{i=1}^n T^*_i$, the union of external knowledge tables
as $ET_n=\bigcup_{i=1}^n E_i$.

At any time $n$, an adversary's explicit background knowledge
consists of $PT_n$ and $ET_n$.
\end{definition}

\begin{definition} [Implicit Background Knowledge]
$\\$ Excluding the explicit background knowledge, the information
which is commonly known to public and can provide help to the
adversary's attack consists of the implicit background knowledge.
Such as the domain and hierarchy of each attribute, the semantic of
each attribute value, the probability of an internal update etc.

%The implicit background knowledge is the knowledge commonly known to
%public. Such as the domain and hierarchy of each attribute, the
%semantic of each attribute value and the probability of each
%internal update.

\end{definition}

Definition~\ref{def_explicit_bk} implies that the adversaries'
explicit background knowledge is incremental and will be enhanced by
each re-publication operation. On the contrary, the implicit
background knowledge is usually static and invariant. At the time of
the $n^{th}$ publication, we denote all the background knowledge of
an adversary as $BK_n$.

\begin{example}
Revisit the example in section~\ref{Invalid-l}.
In the $2^{nd}$ release, the explicit background knowledge includes
$PT_2$ and $ET_2$. $PT_2$ is the union of
table~\ref{Generalization_T1} and table~\ref{Generalization_T2},
$ET_2$ is the union of table~\ref{microdata_T1} and
table~\ref{microdata_T2}, in which the ``Disease'' attribute is
removed.

The rest of information that can also provide assistance to the
adversary consists of the implicit background knowledge. Such as the
domain of work hour per week is $[0, 100]$, the Disease attribute is
categorical etc. Especially, the background knowledge introduced by
internal updates is also implicit: the probability of any attribute
value $a_i$ update to be $a_j$, represented as $P_{trans}(a_i,a_j)$,
is known to public.
\end{example}

Based on the background knowledge, the threat measurement to fully
dynamic dataset, is defined as follows:

\begin{definition}[Disclosure Risk]
For a positive integer $i$, suppose $t_i \in T_i$ holds for $t$.
Before $T^*_{n+1}$ released, the disclosure risk $r_n(t_i)$ is the
\emph{probability} of an adversary~(with the help of $BK_n$) linking
$t_i$ with its actual sensitive value $t_i[S]$.
\end{definition}

As shown below, the disclosure risk in dynamic dataset is also
dynamic. Specifically, it contains two-folded meaning. First, as
time evolves, the disclosure risk of the same version of a record is
fluctuant. Because the dataset re-publications increase the
adversary's explicit background knowledge (definition
\ref{def_explicit_bk}), which may lead to different risk estimation
results at different moments. Second, at the same time, the
disclosure risk of different versions of a record may be different.

\begin{example}
Revisit Julia's records in table~\ref{Generalization_T1}
and~\ref{Generalization_T2}. When $T^*_{1}$ was released, the
disclosure risk of her disease is $50\%$, because the QI-group she
was in has 2 indistinguishable sensitive values.

After the release of $T^*_{2}$, the discourse risk of the disease
she contracted in the $1^{st}$ release increases to be $100\%$.
Meanwhile, now the adversary has different disclosure risks about
Julia's disease: $100\%$ for the old disease in the $1^{st}$ release
and $50\%$ for the new one in the $2^{nd}$ release.

\end{example}

The disclosure risk is a measurement for separate record privacy. In
order to measure the disclosure risk during the entire
re-publication process, we define the \textit{re-publication risk}.

\begin{definition}[Re-publication Risk] Suppose dataset $T$ is fully dynamic. $T^*_{1}, T^*_{2}, T^*_{3},...$
is a sequential release of $T$.

For any integer $i$ and $n$~($0<i\leq n$), if $r_n(t_i)\leq
\alpha$~($\alpha$ is minimum and $\alpha \in [0,1]$) always holds
when $t_i \in T_i$, then we call the re-publication risk of $T$ is
$\alpha$.
\end{definition}

Intuitively, the re-publication risk is a minimized upper-bound of
all disclosure risks. Therefore, we state the problem of
\emph{anonymization of fully dynamic dataset} as, given a fully
dynamic dataset $T$, sequentially release $T^*_{1}, T^*_{2},
T^*_{3},...$ so that the re-publication risk is as lower as possible
and the utility of publications is maximized.
%so that there is no privacy breach and the
%re-publication risk is as lower as possible.

%% file: 3_SUG/SUG.tex
\section{Privacy Disclosure Framework}\label{section_3}
In this section we propose a framework, \emph{Sensitive attribute
Update Graph}~($SUG$), to track an individual's possible sensitive
information and show how the disclosure happens. In
section~\ref{demonstration_app}, we will demonstrate the
applicability of the framework by applying it to the previous work.

\subsection{Sensitive attribute Update
Graph}\label{subsection_sug}

\begin{figure}
 \centering

\begin{minipage}[b]{0.25\textwidth}
\includegraphics[width=1\textwidth]{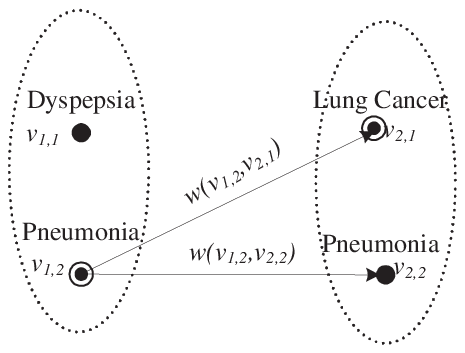}
\end{minipage}
\vspace{-1em}\caption{Julia's $SUG$.Her actual sensitive value is
represented by the circled dot.}\vspace{-1em}\label{Julia-SUG}
\end{figure}

The key idea of $SUG$ is to represent all the possible sensitive
values and updates of a record in a graph: each node represents a
possible sensitive value and each edge represents a feasible update
on a sensitive value. We call an update $U(s_s,s_t)$
\textit{feasible} only if sensitive value $s_s$ has non-zero
probability update to $s_t$.

Suppose that before $T^*_{n+1}$ is released, $t^{'}_{1}, t^{'}_{2},
..., t^{'}_{I}$~($I\leq n$) are the sequential versions of record
$t$ which exist in the corresponding releases of $T$. Formally,
$t$'s $SUG$ is defined as follows:

\begin{definition}[SUG]
Before $T^*_{n+1}$ released, $t$'s sensitive attribute update graph
is denoted by $G_n(V,E)$, such that

\begin{itemize}
\item  there is a \textit{one-to-one} mapping between node
$v_{i,j}$~($v_{i,j}\in V$) and one possible sensitive value
$s_{i,j}$ of $C_i$\footnote{ We call the nodes which represent the
sensitive values in $C_i$ form a corresponding \textit{candidate
node set} $V_i$. }, where $i$ is any integer between $1$ and $I$.

\item the weight of each node $v_{i,j}$ is the probability of
an adversary linking $t^{'}_i$ to $s_{i,j}$ only with the help of
background knowledge ($BK_i-BK_{i-1}$).

\item  an edge $(v_{i,j},v_{i+1,k})$ represents a feasible
update $U(s_{i,j},s_{i+1,k})$ and its weight $w(v_{i,j},v_{i+1,k})$
represents the probability of that feasible update happens, which is
equal to $P_{trans}(s_{i,j},s_{i+1,k})$.
\end{itemize}

\end{definition}

For a $SUG$, the weights of nodes and edges are determined by the
background knowledge. Specifically, the weight of a node is the
linking probability between $t$ and a sensitive candidate value
without the help of historical background knowledge. That's similar
to the linking probability in the publication of static dataset.
However, in dynamic dataset, the linking probability is determined
together with the historical information hidden in the other parts
of the graph.

It is apparent that a record's candidate sensitive sets and their
correlation are all encoded into a $SUG$. From the adversary's
perspective, $G_n(V,E)$ contains all the background knowledge about
$t$'s sensitive information before $T^*_{n+1}$ released. Thus on the
basis of $G_n(V,E)$, s/he can deduce the disclosure risk
$r_n(t^{'}_i)$
 for any integer $i$ between 1 and
$I$.

\begin{example}

Fig.~\ref{Julia-SUG} is Julia's $SUG$ after
Table~\ref{Generalization_T1} and~\ref{Generalization_T2} released.
Without additional knowledge and specific declaration, we follow the
random world assumption~\cite{random-world} that all the sensitive
values in a sensitive candidate set have the equal linking
probability, and the updates on a sensitive value have equal
probability to happen. Thus the weight of every node and edge is 1/2
in fig.~\ref{Julia-SUG}. The disclosure risk $r_2(t^{'}_1)$ is
$100\%$ as $v_{1,1}$ has not outgoing edge; $r_2(t^{'}_2)$ is $50\%$
as both $v_{2,1}$ and $v_{2,2}$ has an incoming edge from $v_{1,2}$.
\end{example}

However, the $SUG$ can be further reduced by excluding some
invalidate nodes and edges. E.g., in fig.~\ref{Julia-SUG}, there's
no edge connect to $v_{1,1}$, that indicates dyspepsia can update to
neither lung cancer nor pneumonia. Thus we know that Julia is
impossible to contract dyspepsia and $v_{1,1}$ has no validate
information. So we can deduce a subgraph only contains the validate
information:

\begin{definition}[feasible sub-$SUG$]
A \emph{feasible sub-$SUG$}\\$G'_n(V',E')$ is a subgraph of
$G_n(V,E)$ induced as follows:

Delete node $v$ and its connected edges from $G_n(V,E)$ if one of
the following conditions holds:
\begin{itemize}
        \item $v\in V_1$ and $deg^{-}(v) =0$;

        \item $v\in V_I$ and $deg^{+}(v) =0$;

        \item $v\in V_i$, $i\in (1, I)$, and at least $deg^{+}(v) =0$ or $deg^{-}(v) =0$ holds;
\end{itemize}
    Repeat the process until no deletion left.

\end{definition}

The above definition also provides a method to induce a feasible
sub-$SUG$ from $SUG$. Actually, the deducing process is also the
major part of the attack: excluding invalidate information so as to
narrow the possible space of the sensitive values.

\begin{figure}

\centering \subfigure[\small A sample $SUG$\label{SUG_1}]{
\begin{minipage}[b]{0.25\textwidth}
\includegraphics[width=1\textwidth]{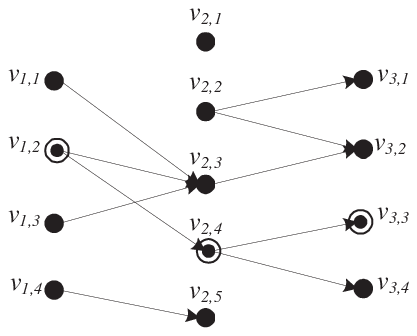}\vspace{-1em}

\end{minipage}

} \subfigure[\small A sample feasible
sub-$SUG$\label{feasible-sub-SUG}]{
\begin{minipage}[b]{0.2\textwidth}

\includegraphics[width=1\textwidth]{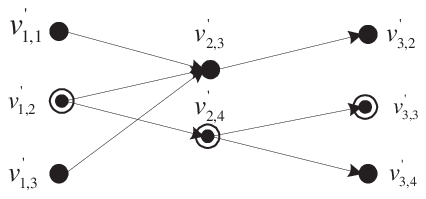}\vspace{-1em}
\end{minipage}
} \caption{\small $SUG$}\vspace{-2em}
\end{figure}

Fig.~\ref{feasible-sub-SUG} is the feasible sub-$SUG$ deduced from
fig.~\ref{SUG_1}. As we observe, in a feasible sub-$SUG$, every path
that begins from node in $V^{'}_1$ and ends with node in
$V^{'}_I$~(we call it a \textit{feasible path}) may represent the
actual path that indicates the evolvement of $t$'s sensitive value.
Thus at a specific time, once we get a record's feasible sub-$SUG$,
we can calculate the probability of each possible path, which can
lead to the estimation of its disclosure risks
$r_n(t^{'}_1),r_n(t^{'}_2),..., r_n(t^{'}_I)$.

\subsection{Disclosure Risk Estimation}

The second part of the attack is the disclosure risk estimation.
Since every path in the feasible sub-$SUG$ may be the one which
contains all the correct sensitive values and updates, the weight
portion of the feasible paths that crossing a node is just the
probability that the correct path contains it, which is also the
probability of linking $t^{'}_i$ to the sensitive value represented
by the node. Thus $r_n(t^{'}_i)$ equals to the weight portion of all
the feasible paths that crossing the node representing $t^{'}_i[S]$
in $V^{'}_i$.

In order to calculate the risk, we first enumerate all the feasible
paths by traversing $G'_n(V^{'},E^{'})$. Then we compute the weight
of every feasible path with the help of related nodes and edges.
Assume
$p_k=\{v^{'}_{1,x_1},v^{'}_{2,x_2},...,v^{'}_{I,x_I}\}$
is any feasible path in $G'_n(V^{'},E^{'})$, $v^{'}_{i,x_i}$ is a
node in $V^{'}_i$. The weight of $p_k$ is the product of all the
nodes and edges it traverses:
\begin{equation}
w(p_k)=w(v^{'}_{I,x_I})\prod
^{I-1}_{i=1}w(v^{'}_{i,x_i})w(v^{'}_{i,x_i},v^{'}_{i+1,x_{i+1}})
\end{equation}

Finally, picking out all the feasible paths that crossing the node
represents $t^{'}_i[S]$, their portion equals to $r_n(t^{'}_i)$:

\begin{equation}\label{eq3}
r_n(t^{'}_i)=\frac{\sum^{K_i}_{k^{'}=1}w(p_{k^{'}})}{\sum^{K}_{k=1}w(p_{k})}
\end{equation}
$K$ is the total number of feasible paths and $K_i$ is the count of
feasible paths that crossing the node represents $t^{'}_i[S]$.

\begin{example}
Consider the feasible sub-$SUG$ in fig.~\ref{feasible-sub-SUG}.
There are totally 5 feasible paths in this graph. Enumerating them
from top to down, their weights are 1/18, 1/36, 1/72, 1/72 and 1/18
respectively. The sum is 1/6.

There are 3 feasible paths crossing node $v^{'}_{1,2}$:
$\{v^{'}_{1,2}, v^{'}_{2,3}, v^{'}_{3,2}\}$, $\{v^{'}_{1,2},
v^{'}_{2,4}, v^{'}_{3,3}\}$ and $\{v^{'}_{1,2}, v^{'}_{2,4},
v^{'}_{3,4}\}$. So $r_3(t^{'}_1)=(1/36+1/72+1/72)/(1/6)=1/3$.
Similarly, we have $r_3(t^{'}_2)=1/6$ and $r_3(t^{'}_3)=1/12$.

\end{example}

\newtheorem{lemma}{Lemma}

We can estimate $r_n(t^{'}_1), r_n(t^{'}_2),..., r_n(t^{'}_I)$ based
on $G^{'}_n(V^{'}, E^{'})$. Generally, for any positive integer $j$,
in order to estimate $r_j(t^{'}_1), r_j(t^{'}_2),..., r_j(t^{'}_I)$,
we should construct its feasible sub-$SUG$ $G'_j(V^{'},E^{'})$ with
the help of $BK_j$.

The increasing releases of $T$ will lead $t$'s feasible sub-$SUG$
dynamic and growing. Thus $r_j(t^{'}_i)$ is usually not equal to
$r_k(t^{'}_i)$~($j\neq k$). Because in $t$'s different feasible
sub-$SUG$s, the weight portion of feasible paths that crossing the
same node is usually variant. However, there are still exceptions:
\begin{lemma}\label{corollary}
If all the sensitive values in the sensitive domain can randomly
updated to any other value~(including itself), the disclosure risks
of the existing sensitive information are invariant regardless how
to release the new publications.
\end{lemma}
\begin{proof}
The proof of lemmas can be found in~\cite{online-version}.
\end{proof}

The lemma also implies the anonymization problem of dynamic dataset
can be reduced to several independent anonymization problem of
static dataset when the internal updates on sensitive values are
totally random. Because the random updates of sensitive values lead
different publications to no correlation: $T_{n}$ and $T_{n+1}$ are
entirely independent.

Generally, we have the
following lemma for dynamic dataset which theoretically demonstrates
at what time the disclosure of sensitive information happens:

\begin{lemma}\label{v=1}
Regardless how to publish the dataset, for any positive integer $i$
and $n$~($i\leq n$), $r_n(t^{'}_i)=1$ holds iff $|V^{'}_i|=1$ holds
in the corresponding feasible sub-$SUG$.
\end{lemma}\vspace{-1em}

\subsection{$SUG$ Applicability Demonstration}\label{demonstration_app}
As mentioned, $SUG$ is a general privacy disclosure framework for
re-publication problem. Exploiting it to analysis any re-publication
problem, we follow two steps:
\begin{enumerate}
\item Constructing the record's $SUG$ and deducing the corresponding
feasible sub-$SUG$;

\item Calculating the weight of each feasible path and estimating
disclosure risks.
\end{enumerate}

Let us apply the framework to re-publish external dynamic
dataset~\cite{xiao:m-invariance}. Since there is no internal update
in the external dynamic dataset, each node have at most one incoming
edge and one outgoing edge in a record's $SUG$; each edge connects
two nodes which represent the same sensitive value. After excluding
the invalidate information in the $SUG$, the feasible sub-$SUG$ must
have the following characteristics:

\begin{enumerate}
\item For each feasible path, all the nodes it crossed represent the same sensitive
value;
\item For each node, there is only one feasible path crossing it.
\end{enumerate}

Intuitively, the feasible sub-$SUG$ contains several parallel
feasible paths and each one contains the same sensitive value.
According to lemma~\ref{v=1}, the disclosure will occur when there
is only one feasible path left. The analysis also hints us that, if
we can guarantee that the feasible sub-$SUG$ always have several
indistinguishable feasible paths, the disclosure will not
happen\footnote{In fact, that is the basic idea of
\textit{m}-Invariance: it guarantees that there are always
\textit{m} parallel feasible paths for each record.}. Moreover,
employing our estimation method, we will get that the re-publication
risk of \textit{m}-Invariance is $1/m$ as each feasible path has
equal weight.

The analysis above also convinced us that re-publication of external
dynamic dataset is a special case of our problem. Revisiting fully
dynamic dataset, as illustrated in fig.~\ref{feasible-sub-SUG}, the
$SUG$ of each record is more complex and the risks are difficult to
control.
However, if we can make a similar guarantee: in each record's
feasible sub-$SUG$, there always exists several indistinguishable
feasible paths and each candidate node set contains several nodes,
at least the disclosure will not occur.

%% file: 4_principle/principle.tex
\section{Anonymization Principle}\label{section_4}
According to the analysis in the previous section, if a record's
sensitive information is well protected in each separate
publication, the disclosure is mainly rose by the pruning to its
$SUG$. In other words, if we prevent the possible pruning to the
record's $SUG$ and always guarantee $|V^{'}_i|>1$ for all $i$, the
disclosure of sensitive information will never happen.

Referring to a dataset containing a mount of records, we should pay
more attention: when publishing the dataset, we need guarantee that
there will be no pruning to \textit{all} the records at \textit{any}
time, as to prevent the chain-actions of
disclosure~\cite{corruption}.

Specifically, two requirements need to be met:

\begin{enumerate}
\item All the records' sensitive information is well preserved in
each separate publication;
\item At any time, there is no pruning to all the records' $SUG$ so as to maintain the indistinguishability of sensitive values.
\end{enumerate}

In this paper, we use \textit{m}-unique~\cite{xiao:m-invariance} to
illustrate the sensitive value indistinguishability: if there are at
least \textit{m} records in $QI$-group $g$ and all of them have
distinct sensitive values, we call $g$ is \textit{m}-unique; a
published table is \textit{m}-unique if all the $QI$-groups in it
are \textit{m}-unique.

\subsection{\textit{m}-Distinct}

Before presenting our method, we formulate the following concept to
describe the update candidates of a value:

\begin{definition}[Candidate Update Set]
Suppose $a$ is an element in the domain of attribute
$\mathcal{A}$~($a\in dom(\mathcal{A})$), its \emph{candidate update
set} $CUS(a)$ is the union of some elements in $dom(\mathcal{A})$,
such that $a$ has \textit{non-zero} update probability to it.
\end{definition}

Note that if $b \in CUS(a)$, then $CUS(b) \subseteq CUS(a)$ must
hold\footnote{The result is straightforward using the method of
Reduction to Absurdity.}.
%according to the transitivity of updates.
Similarly, we have the
following notion for a group of sensitive values:

\begin{definition}[Update Set Signature]
Suppose $QI$-group $g$ contains $n$ records and their sensitive
values are $s_1,s_2,...,s_n$, respectively. Then $g$'s
\textit{update set signature} $USS(g)$ is a multi-set:
%$\bigcup_{i=1}^{n}CUS(s_i)$.
\{$CUS(s_1),CUS(s_2), ... ,CUS(s_n)$\}.

\end{definition}

Since $USS$ is a multi-set of $CUS$, the same $CUS$ may appear
several times in a $USS$, because several records may have the same
sensitive value and different sensitive values may even have the
equal candidate update set. Record $t$'s update set signature, which
is inherited from the $QI$-group it is in, is denoted by $USS(t)$.
It is obvious that a record's $USS$ is dynamic as the re-publication
progress evolves, because in different time the sensitive values of
its $QI$-group are variant.

In this paper, we say that $USS_i$ and $USS_j$~($i\not=j$) are
\textit{intersectable}, if they have equal number of $CUS$ and there
exists a one-to-one map between two $CUS$ in $USS_i$ and $USS_j$,
such that the intersection of the two $CUS$ is non-empty; moreover,
if the $CUS$ of $USS_j$ is a subset of the $CUS$ of $USS_i$, we call
$USS_i$ \textit{implies} $USS_j$~(denote as $USS_i\supseteq USS_j$).

Next, we explains under what conditions, a set of values is a legal
update instances of a $USS$:

\begin{definition}[Legal Update Instance]\label{illegal}

A set of sensitive values $S=\{s_1, s_2, ..., s_n\}$ is a
\textit{legal update instance} of a $USS$ if the following
conditions hold:
\begin{enumerate}
\item The number of sensitive values in $S$ equals to the number of
$CUS$ in the $USS$: $|S|=|USS|$.

\item For any value $s_i$ in $S$, there is at least one candidate
update set $CUS_j$ such that $s_i\in CUS_j$.

\item For any candidate update set $CUS_j$ in $USS$, there is at least
one sensitive value $s_i$ in $S$ such that $s_i\in CUS_j$.
\end{enumerate}

\end{definition}

If a group of sensitive values are a legal update instance of a
$USS$, in the perspective of adversary, every value in it can not be
excluded. Suppose $t$'s candidate sensitive set $C$ is a legal
update instance of its $USS$ in the previous publication, then the
deduce procedure~(as illustrated in section~\ref{subsection_sug})
can not exclude any node or edge: all the information in its $SUG$
are validate. Hence the threats rose by invalidate information
exclusion are prevented.

\begin{example}\label{illegal-ex}
In the example of section~\ref{motiv-distinct}, Julia's candidate
sensitive set $C_1$ is $\{Dyspepsia, Pneumonia\}$. With the help of
implicit background knowledge, we know
that\\$CUS(Dyspepsia)=\{Dyspepsia,
Gastritis,other\;digestive\;\\system\;diseases\}$ and
$CUS(Pneumonia)=\\\{Pneumonia,Flu,Lung\;Cancer,
other\;respiratory\;\\system\;diseases\}$. Thus Julia's update set
signature in the $1^{st}$ release is $\{CUS(Dyspepsia),
CUS(Pneumonia)\}$.

According to definition~\ref{illegal}, if we randomly pick out an
element from $CUS(Dyspepsia)$ and $CUS(Pneumonia)$ respectively,
then the two elements must be a legal instance of $USS(Julia_1)$.

\end{example}

Now we are ready for our anonymization principle:

\begin{definition}[\textit{m}-Distinct]\label{m-distinct}
$T$ is a dynamic dataset, a sequential releases of $T$: $T^*_{1},
T^*_{2}$, $T^*_{3}$, $...$, $T^*_{n}$ are \textit{m}-Distinct if it
meets:%
\begin{enumerate}
\item     For all $i\in[1,n]$, $T^{*}_i$ is \textit{m}-unique.

\item     Suppose for any record $t$, $T_i$ and $T_j$~($i < j$) are
two neighboring  releases which both contain $t$~($t_i\in T_i$,
$t_j\in T_j$). For all $i\in[1,n]$, $t_j$'s candidate sensitive set
$C_j$ is a legal update instance of $USS(t_i)$.
\end{enumerate}

\end{definition}

The rationale of \textit{m}-Distinct is that, we adopt
\textit{m}-unique to maintain the indistinguishability of sensitive
values in each separate publication; then when releasing new
publication, we carefully partition the records so that the
indistinguishability of sensitive values is still maintained. In
other words, the concept of ``legal update instance'' guarantees
that there is no inference rose by information exclusion.

Revisit example~\ref{illegal-ex}, since Julia's candidate sensitive
set $C_2$=$\{Dyspepsia, Lung\;Cancer\}$ is a legal update instance
of $\{CUS(Dyspepsia), CUS(Pneumonia)\}$, $T^{*}_1$ and $T^{*}_2$ are
\textit{2}-Distinct with respect to Julia.

Deriving from definition~\ref{m-distinct}, when releasing a new
version of $T$ and maintaining the \textit{m}-Distinct property
meanwhile, we only need the information of most recent versions of
the records. Specifically, if the two following conditions hold:

\begin{enumerate}

\item the new version $T^{*}_{new}$ is \textit{m}-unique;

\item for any record $t$~($t_{new}\in T_{new}$), suppose $t_{pre}$ is $t$'s most recent version, then $t_{new}$'s candidate sensitive set is a legal update instance of
$USS(t_{pre})$.
\end{enumerate}

Then the sequential release including $T^{*}_{new}$ are also
\textit{m}-Distinct. Furthermore, we have the following lemma:

\begin{lemma}\label{v>=m}

If a sequential releases of $T$: $T^*_{1}, T^*_{2}$, $T^*_{3}$,
$...$, $T^*_{n}$ are \textit{m}-Distinct, then for any record $t\in
T$, $|V^{'}_i|\geq m$ holds for all the candidate node sets in its
feasible sub-$SUG$ $G'_n(V',E')$.

\end{lemma}

Lemma~\ref{v>=m} reveals that the disclosure will not occur if the
releases are \textit{m}-Distinct. A larger $m$ usually makes the
disclosure more difficult because more values are indistinguishable
in each QI-group.

\subsection{\textit{m}-Distinct Extension}
\textit{m}-Distinct guarantees that no disclosure of sensitive
values will occur, however, sometimes more strict anonymization
principle may be needed to limit the re-publication risk. Thus we
have the following principle called \textit{m-Distinct$^{*}$}:

\begin{enumerate}
\item the requirements of \textit{m}-Distinct hold.

\item Suppose for any record $t$, $T_{first}$ is the first release
contains $t$~($t_{first}\in T_{first}$). then $CUS_{\alpha} \cap
CUS_{\beta}=\phi$~($\alpha \neq \beta$) holds for any two candidate
update sets in $USS(t_{first})$.

\end{enumerate}

Then the following consequence holds:

\begin{lemma}\label{risk<=1/m}

If a sequential releases of $T$: $T^*_{1}, T^*_{2}$, $T^*_{3}$,
$...$, $T^*_{n}$ are \textit{m}-Distinct$^{*}$, then the
re-publication risk is at most $1/m$.

\end{lemma}

The key of \textit{m}-Distinct$^{*}$ is the $2^{nd}$ condition. It
implies that, in the same $QI$-group, every sensitive value's $CUS$
does not overlap with the other values'.

By applying our privacy disclosure framework,
\textit{m}-Distinct$^{*}$ can limit the re-publication risk to $1/m$
because it guarantees that there are at least $m$ parallel feasible
paths in every record's feasible sub-$SUG$. However,
\textit{m}-Distinct$^{*}$ may not be met in general case: there may
not exist $m$ sensitive values to form a QI-group in which there is
no overlap between the $CUS$ of any two sensitive values.

%% file: 5_algorithm/algorithm.tex
\section{Algorithm}\label{section_5}

We now present an algorithm to meet \textit{m}-Distinct. According
to the analysis in previous section, if every new release of the
dataset meets the two conditions in previous section, then
\textit{m}-Distinct persists in the sequential release. Thus we put
the attention on releasing $T^{*}_n$ based on the previous releases.

When anonymizing $T_n$, the crucial part of maintaining
\textit{m}-Distinct property is that, every record's new candidate
sensitive set should be a legal update instance of its previous
$USS$. The basic idea of our algorithm is to assign records to
proper bucket according to their $USS$ such that we can always find
a way to partition records into $QI$-group, of which the candidate
sensitive set is a legal update instance of these records. The
overview of our algorithm is described in
Algorithm~\ref{General_Procedure_of_Our_Algorithm}.

\begin{algorithm}
\small \caption{Overview of Our
Algorithm}\label{General_Procedure_of_Our_Algorithm}
\begin{algorithmic}[1]

\REQUIRE  $T_n$: the $n^{th}$ version of $T$;

$m$: the user configured parameter for \textit{m}-Distinct;

$t_{pre}$: the most recent version of $t$, where $t\in T_n$ and $t$
has appeared in $T$ before;

\STATE \textit{create} buckets $Q_{buc}$ according to the records
which has $t_{pre}$;

\STATE \textit{assign} records to proper position of proper bucket
in $Q_{buc}$;

\STATE \textit{partition} the unassigned records into
\textit{m}-unique $QI$-groups;

\STATE recursively \textit{split} each bucket into two until no more
split left~(forms $QI$-groups);

\STATE \textit{generalize} each $QI$-group;

\STATE \textit{publish} the generalized $QI$-groups and counterfeit
statistics.
\end{algorithmic}
\end{algorithm}

In our algorithm we introduce counterfeit records when no enough
record in the dataset helps to meet \textit{m}-Distinct.
Note that the only usage of the counterfeit records is to maintain
the sensitive value indistinguishability of a $QI$-group.

Except for meeting the anonymization principle, we also aims to
minimize two criterions: the number of counterfeit records and the
generalization of $QI$ attributes. Because more counterfeit records
will cover up more characteristics of the original dataset and more
generalization will lead to more information loss, both are harm to
the dataset utility.

Our algorithm mainly contains the three following phases.

\subsection{Phase 1: Creating Buckets}
\vspace{-1em}

\begin{algorithm}
\small

\caption{Create Buckets}\label{creating}

\begin{algorithmic}[1]

\REQUIRE $Q_{rec}$: the queue of records in $T_n$;

%any record $t$'s most recent version $t_{pre}$ if it exists.
$t_{pre}$: the most recent version of $t$, where $t\in T_n$ and $t$
has appeared in $T$ before;

\STATE Let $Q_{buc}$ and $Q_{tmp}$ be empty queues of buckets;

\FORALL{record $t$ in $Q_{rec}$}\label{step-1-b}

\IF{$t_{pre}$ exists}

\STATE  create $B_{new}$ such that $USS(B_{new})=USS(t_{pre})$;

\IF{$B_{new}$ does not exist in $Q_{buc}$}

\STATE \textit{enqueue}($Q_{buc}, B_{new}$);

\ENDIF

\ENDIF

\ENDFOR\label{step-1-e}

\FOR{every 2 buckets $B_i, B_j\in Q_{buc}(i<j)$}\label{step-2-b}

\IF{$USS(B_i)$ and $USS(B_j)$ are \textit{intersectable}}

\STATE pick out their highest scored intersection plan $USS_{new}$;

\STATE create $B_{new}$ such that $USS(B_{new})=USS_{new}$;

\IF{$B_{new}$ does not exist in $Q_{buc}$ and $Q_{tmp}$}

\STATE \textit{enqueue}($Q_{tmp}, B_{new}$);

\ENDIF

\ENDIF

\ENDFOR\label{step-2-e}

\STATE \textit{append}($Q_{buc}, Q_{tmp}$);

\STATE return $Q_{buc}$;

\end{algorithmic}
\end{algorithm}
\vspace{-0.5em}

Our algorithm~(algorithm~\ref{creating}) will first create buckets
which the records are possibly in. Note that a bucket is only
identified by its $USS$.

Suppose for any record $t\in T_n$, $t_{pre}$ is $t$'s most recent
version. First, we create a bucket $B_{new}$ for $t$, such that
$USS(B_{new})$ equals to
$USS(t_{pre})$~(lines~\ref{step-1-b}-\ref{step-1-e}). We only skip
record $t$ if such bucket is already exist, or this is the first
time $t$ appears in the dataset. We denote an \emph{entry} of bucket
$B$ as a candidate update set in $USS(B)$. The number of entries
equals to the number of candidate update sets of $B$.

In the second step~(lines~\ref{step-2-b}-\ref{step-2-e}), we
generate new bucket based on the buckets created in the previous
step: if any two buckets are \textit{intersectable}, we create a new
bucket whose update set signature is the intersection of their
$USS$; if there are several possible intersection plans, we choose
the one with highest score: the higher proportion of the overlapped
elements, the higher score of the intersection plan.

At the end of this phase, we have all the possible buckets for the
records of $T_n$ which have appeared before. For the new records, we
will create new bucket for them later, if no existing bucket is
suitable.

\subsection{Phase 2: Assigning Records}

\begin{algorithm}
\small \caption{Assign Records}
\begin{algorithmic}[1]

\REQUIRE $Q_{rec}$: the queue of sorted records~($CNT_{buc}\geq 1$
for all);

$CNT_{buc}(t)$: the number of buckets $t$ can be assigned;

$Q_{t}$: the queue of suitable buckets for record $t$.

\WHILE{$Q_{rec}$ is not empty}

\STATE $t\leftarrow$\textit{dequeue}($Q_{rec}$);

\STATE max\_score$\leftarrow -\infty$;\COMMENT{global maximum score
of $t$}

\WHILE{$Q_{t}$ is not empty}\label{assign-possible-bucket}

\STATE $B\leftarrow$\textit{dequeue}($Q_{t}$);

\STATE buc\_score$\leftarrow -\infty$;\COMMENT{maximum score of
$t$'s assignment in $B$}

\FOR{entry $e_i$ in $B$}

\IF{$t[S]\in e_i$}\label{assign-possible-entry}

\STATE tmp\_score$\leftarrow$\textit{get\_score}($t, B,
e_i$);\label{assign-get-score}

\IF{tmp\_score==buc\_score}

\STATE buc\_entry$\leftarrow$\textit{choose\_entry}(buc\_entry,
$i$);\label{assign-choose-entry}

\ELSIF{tmp\_score$>$buc\_score}

\STATE buc\_score$\leftarrow$tmp\_score;

\STATE buc\_entry$\leftarrow$$i$;

\ENDIF

\ENDIF

\ENDFOR

\IF{buc\_score$>$max\_score}

\STATE max\_score$\leftarrow$buc\_score;

\STATE max\_entry$\leftarrow$buc\_entry;

\STATE max\_buc$\leftarrow B$;

\ENDIF

\ENDWHILE

\STATE \textit{assign}($t$,max\_buc,
max\_entry);\label{assign-assign-record}

\ENDWHILE

\end{algorithmic}
\end{algorithm}

The main task of this phase is to assign records to proper bucket
and the corresponding entry. If record $t$ and bucket $B$ meet
$USS(t_{pre})\supseteq USS(B)$ and $t[S]$ is covered by $USS(B)$,
then $t$ can be assigned to $B$. The reason is that, when we pick
out an record from each entry of the bucket and forms a
$QI$-group~(will carry on in next phase), if there is no duplicate
sensitive value in the group, the $QI$-group must hold
\textit{m}-Distinct. Because the candidate sensitive set of $t$ now
must be a legal update instance of $USS(B)$ as well as
$USS(t_{pre})$.

Referring to a record which appears in the dataset the first time,
it can be assign to a bucket only if its current sensitive value is
covered by the bucket's $USS$.

In order to facilitate the task, we first calculate $CNT_{buc}(t)$,
the number of buckets $t$ can be assigned, and sort the records
increasingly according to their $CNT_{buc}$.

The records which have no existing buckets to be assigned
in~($CNT_{buc}$=0), must also be the first time appears in the
dataset. Thus we process them separately: partitioning them into
$QI$-groups which are \textit{m}-unique. It can be done by
exploiting existing anonymization
algorithms~\cite{mondrian,l-diversity,xiao:m-invariance} because
they have no previous version involved. Note that counterfeit
records will be added in case they are not
\textit{m}-eligible\footnote{a group of records are
\textit{m}-eligible~\cite{xiao:m-invariance}, if there are no more
than $1/m$ records have the same sensitive value. These records can
be partitioned into \textit{m}-unique $QI$-groups only if they are
\textit{m}-eligible~\cite{l-diversity}.}.

The rest of the records, which can be assigned to at least one
bucket, will be assigned to a bucket sequentially as algorithm 3.
Since there are $K!$~(denote $K$ as the number of rest records)
orders to assign these records, our greed algorithm starts the
assignment by processing the records with least $CNT_{buc}$, because
they have less optional buckets and can be determined with less
overheads.

In algorithm 3, we consider each possible
bucket~(line~\ref{assign-possible-bucket}) and entry for a record so
as to get the highest scored assignment. A record $t$ can be
assigned to an entry $e_i$ only if the $CUS$ it represented contains
$t[S]$~(line~\ref{assign-possible-entry}).
Specifically, the score of $t$ with respect to $e_i$ of
$B$~(line~\ref{assign-get-score}) is calculated as follows:

(i) We define $\epsilon$ to indicate $t$'s contribution to
counterfeit counts if $t$ is assigned to $e_i$. $\epsilon=1$ means
$t$'s assignment will not increase the counterfeit count in $B$;
otherwise, $\epsilon$ is $-1$. So we first calculate parameter
$\delta$, such that
$\delta=max\{F_{max},|e_1|,|e_2|,...,|e_{|USS(B)|}|\}$, where
$F_{max}$ is the maximal frequency of sensitive value in $B$. Then
we set $\epsilon$ to be $-1$ if the frequency of $t[S]$ in $B$
equals $\delta$ or $|e_i|$ equals $\delta$.

(ii) We also define a parameter to indicate $t$'s contribution to
the further generation: $\lambda =Z_{aft}/Z_{bef}$. $Z_{bef}$ and
$Z_{aft}$ are the $|QI|$-dimensional area generated by all the
records in $B$ before and after $t$'s assignment. Apparently,
$\lambda\geq 1$ and a larger value indicates $t$'s assignment brings
into more generation.

(iii) At last we return the following score:
\begin{equation}
score= \left\{
\begin{array}{ll}
1/\lambda & if\; \epsilon=1 \\
-\lambda & if\; \epsilon=-1
\end{array}
\right.
\end{equation}

The above equation blends $t$'s contributions to counterfeit count
and generalization together.
Obviously, a larger score of $t$'s assignment shows that it will
bring into less counterfeit records and generalization.

When two entries in $B$ has the same score, we assign the record
into the entry with less already assigned
records~(line~\ref{assign-choose-entry} \textit{choose\_entry}), so
as to get a more balanced bucket. Once we get the assignment with
least score, we immediately assign this record to the bucket by
pushing it into the corresponding bucket
entry~(line~\ref{assign-assign-record}).

After all the records are assigned, the buckets will be balanced
with counterfeit records so as they are \textit{m}-eligible. Thus we
calculate $\delta$ as before, then add counterfeit records to the
entries so that each entry has $\delta$ records.

\subsection{Phase 3: Generating $QI$-groups}\label{phase_3}

Now every bucket is \textit{m}-eligible and is well prepared for
generating $QI$-groups. Since there are $|USS(B)|$ entries in $B$
and each one has $\delta$ records, it is workable to split the
bucket into $\delta$ $QI$-groups: each one contains only one record
of an entry and all the records have distinct sensitive values.

In this phase, we will recursively split each bucket into two
children until only one record left in each entry. In order to
perform further split on the generated buckets, the child buckets
should also be \textit{m}-eligible. Specifically, both of them
should meet the following conditions:

\begin{enumerate}
\item balanced;
\item $F_{max}$
should not be larger than the number of records in each
entry.
\end{enumerate}

Besides, in order to generate $QI$-groups with least information
loss, each split we aims to minimize the generalization. Similar
to~\cite{xiao:m-invariance}, we calculate score for a split plan as
follows:

\begin{equation}
split\_score=\sum^{2}_{i=1}(|B_i|\cdot\sum^{d}_{j=1}\frac{l_{i,j}}{I_j})
\end{equation}
where $l_{i,j}$ and $I_j$ are the minimum interval of attribute
$q_j$ in $B_i$ and $B$, respectively.

To find the split plan with least score, we organize the records of
a bucket in a queue and sort them according to attribute $q_i$. Then
we greedily pick out records from the queue so as to form two child
buckets. The split plan with least score is kept. After we do the
above procedure for all the $QI$ attributes, we choose the minimum
one and apply it to $B$.

The key of this phase is how to pick out records so as to form two
child buckets which both meet the above conditions. In our
algorithm, each time we traversal the queue and pick out $|USS(B)|$
records, which are all from different entries and have no duplicate
sensitive value. The pick-out procedure executes recursively so as
to pick out more records and generate all the possible split plans.
In the worst case, ${\delta^{'}}^{|USS(B)|-1}$ operations may be
performed in order to pick out $|USS(B)|$ legal records.
$\delta^{'}$ is the current number of records which are still left
in each entry.

Note that if $B$ can not be split again and it has counterfeit
record in entry $e_i$, we will randomly assign the counterfeit
record a sensitive value, which is pick out from the $CUS$ that
$e_i$ represented and different from the existing values in $B$.

At last, we generalize all the $QI$-groups formed in phase 2 and 3
and publish them together with the corresponding counterfeit
statistics.

\subsection{Extension}
The presented algorithm is a general method to meet
\textit{m}-Distinct. According to the definition of
\textit{m}-Distinct$^{*}$, it has an additional constrain on the
$1^{st}$ release of any record in contrast to \textit{m}-Distinct.
Thus we only need to handle the $QI$-group with new records
particularly: grantees that it has at least \textit{m} records with
different sensitive values and the $CUS$ of any two records'
sensitive values are not overlapped.

To achieving this, in each publication, we will first check whether
there are suitable buckets created in phase 1 for the new record.
The record will be assigned in if such a bucket found, otherwise we
partitioning these new record into new $QI$-groups as did in the
anonymization of static dataset but with an addition criterion: in
each $QI$-group, no overlap exists between any two sensitive values'
$CUS$. For the old records, we process them no difference to the
procedure in \textit{m}-Distinct.

%% file: 6_experiments/experiments.tex
\section{Experiments}\label{section_6}
The experiments were performed on a 3GHz Intel IV processor machine
with 2GB memory. All the algorithms are implemented in C++.

\subsection{Experiment Setup}
We use a real dataset \textit{OCC} from \textit{http://ipums.org},
which is also adopted by~\cite{anatomy,xiao:m-invariance}. The
dataset consists of 200k records with four $QI$ attributes and one
sensitive attribute. More detail information of the dataset is given
in table~\ref{occ-desc}. Vital parameters of the experiment are set
as follows:

\textbf{External Update.} Since the external update property is well
investigated in~\cite{xiao:m-invariance}, in our experiment, we use
a fixed \textit{external update rate}: we began to publish $T_1$
with \textbf{20,000} records which are randomly chose from the
original dataset; then in each new release $T_i$, we randomly remove
\textbf{2,000} records from $T_{i-1}$ and insert \textbf{5,000}
records from the rest of records. The dataset will be re-published
20 times.

\textbf{Internal Update.} As no existing dataset contains internal
updates information explicitly, we generate internal updates
according to the semantic of each attribute. In the span of $[i,
i+1]$, the internal updates are configured as follows:

\begin{itemize}
\item \textit{Age}: the age of each record will increase 1 till reach 100;

\item \textit{Gender}: will not change;

\item \textit{Marital Status/ Education}:
will update according to the specific semantic of each value. E.g.,
the marital status of a record may update from \textit{married} to
any one in \textit{\{married, divorced, separated, widowed\}} but
can not update to be \textit{never-married}; its education may
update from \textit{bachelor} to any eduction not lower than
bachelor.

\item \textit{Occupation}: since the internal update on sensitive
attribute is critical to the problem of this paper, we introduce
\textit{internal update diameter \textbf{d}} to describe the
flexibility of internal updates on sensitive attribute. An sensitive
value's \textit{d} equals to the size of its candidate
set\footnote{For the convenience, we set all the sensitive values'
candidate sets to be the same size in our experiment.}. Apparently,
a large diameter indicates more flexible internal updates.

By default, we set $d$ to be 10, which means
a person's occupation may stay the same or change to be nine other
similar jobs with equal probabilities. Noticing that in each
publication, we set the internal updates on sensitive values only
randomly occurred on \textbf{5,000} records.

\end{itemize}

\begin{table}
\centering \small{

\caption{Dataset Description}\label{occ-desc}\vspace{-1em}
\begin{tabular}{|l|c|c|c|c|c|}
\hline
 \bf{attribute}&Age& Gender &Marital.& Education & Occupation\\\hline

\bf{dom. size}&100&2&6&17&50\\\hline

\bf{type} &num.&cat.&cat.&cat.&cat.\\\hline

\end{tabular}
\vspace{-2em} }

\end{table}

\subsection{Invalidation of Existing methods}
We first perform experiments to show the inadequacy of the existing
anonymization methods.

\begin{figure}

\centering \subfigure[\small{Vs.
time}\label{dis_totalSens_time_Vs_L}]{
\begin{minipage}[b]{0.22\textwidth}
\includegraphics[width=1\textwidth]{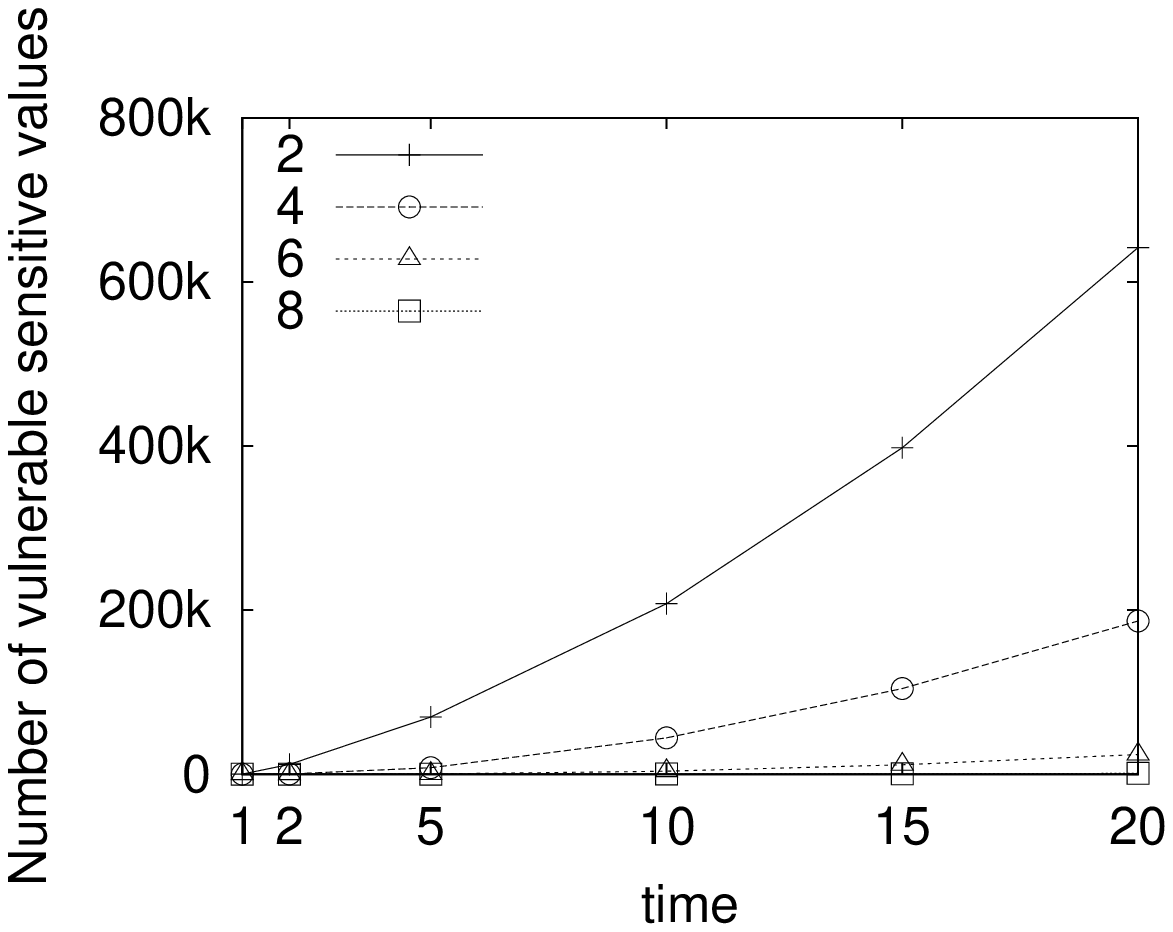}

\end{minipage}%
} \subfigure[\small{Vs. \textit{d}}\label{dis_totalSens_d_Vs_L}]{
\begin{minipage}[b]{0.22\textwidth}

\includegraphics[width=1\textwidth]{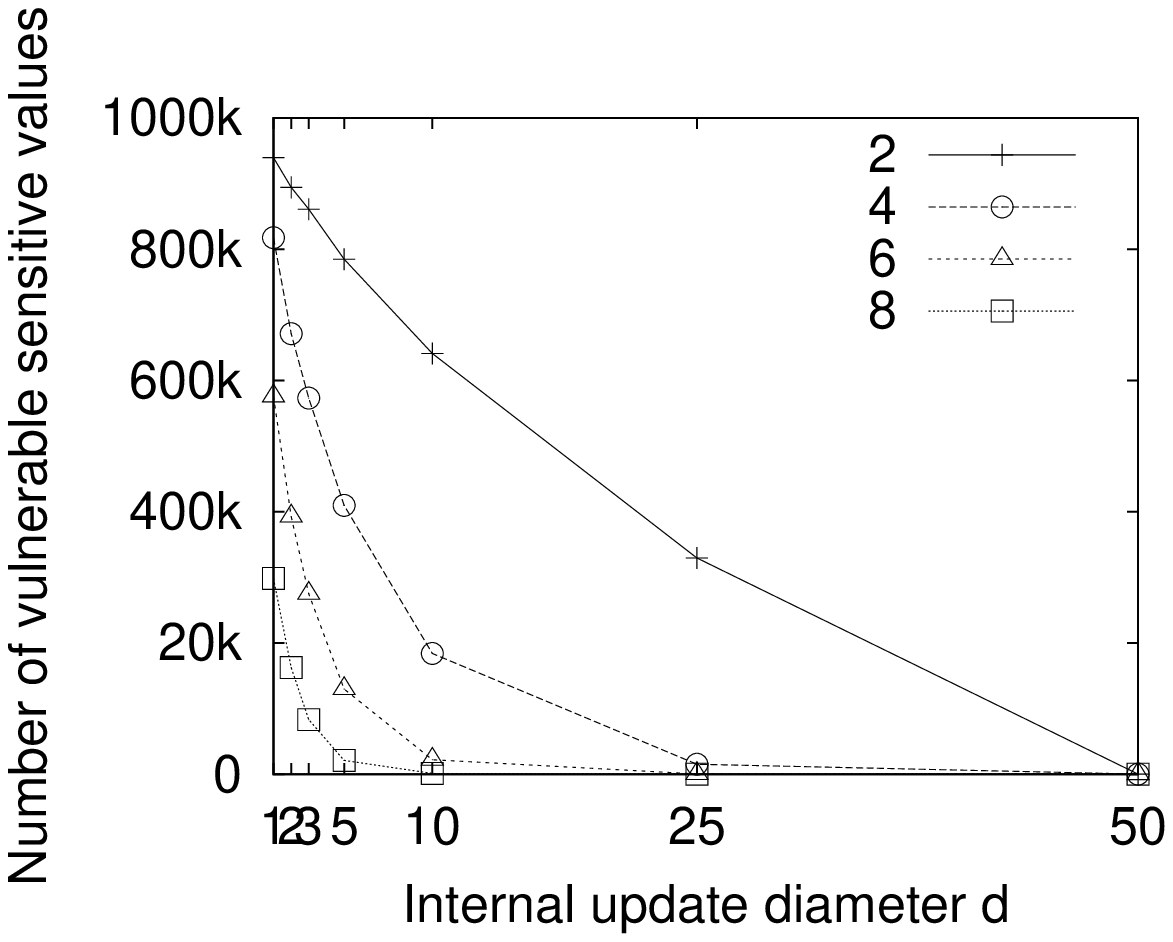}
\end{minipage}

} \vspace{-1em}\caption{Total vulnerable counts}\vspace{-1em}
\end{figure}

The \textit{l}-diversity algorithm in~\cite{l-diversity} is
exploited to re-publish the dynamic dataset.
Fig.~\ref{dis_totalSens_time_Vs_L} demonstrates the total number of
vulnerable sensitive values if the dataset is re-published using
\textit{l}-diversity. Note that there are $n$ versions of sensitive
value related to a record if it is published $n$ times.
The result confirms that \textit{l}-diversity is insufficient to
re-publish dynamic dataset,
e.g. about $66\%$ percentage of the published sensitive values are
disclosed when the dataset is published 20 times using
$2$-diversity. Although a larger $l$ will lead less disclosure, the
number of disclosed sensitive values increases as the re-publication
process evolves.

We also test the number of vulnerable sensitive values versus
internal update diameter~(fig.~\ref{dis_totalSens_d_Vs_L}). A
smaller diameter usually leads to more vulnerable sensitive
information, because the updates on sensitive values are less
flexible and more pruning will be performed when deducing the
feasible sub-$SUG$.
Besides, the total vulnerable information decreases as $l$ grows,
because lemma~\ref{v=1} is more difficult to be met when more
records are plunged into a group.

\begin{figure}

\centering \subfigure[\small{Vs.
time}\label{invalidate_totalTuples_time_Vs_m}]{
\begin{minipage}[b]{0.22\textwidth}
\includegraphics[width=1\textwidth]{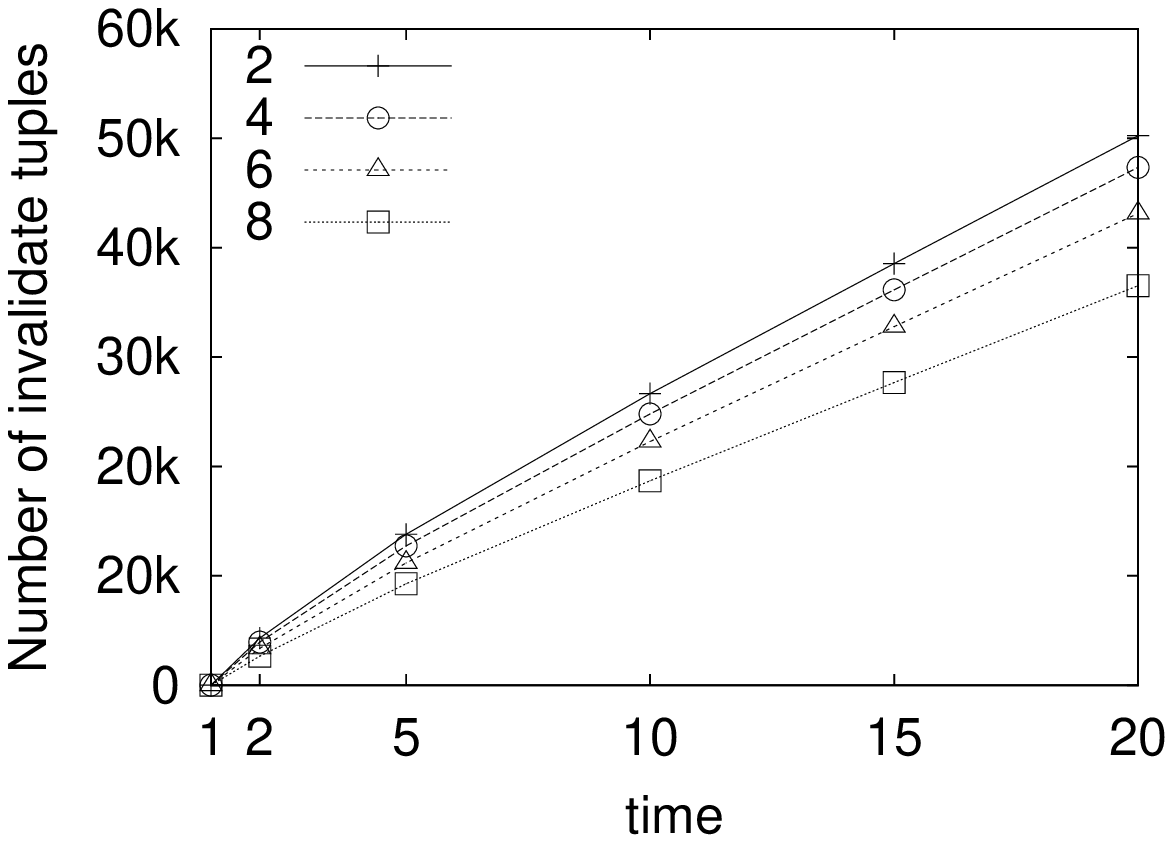}

\end{minipage}%
} \subfigure[\small{Vs.
\textit{d}}\label{invalidate_totalTuples_d_Vs_m}]{
\begin{minipage}[b]{0.22\textwidth}

\includegraphics[width=1\textwidth]{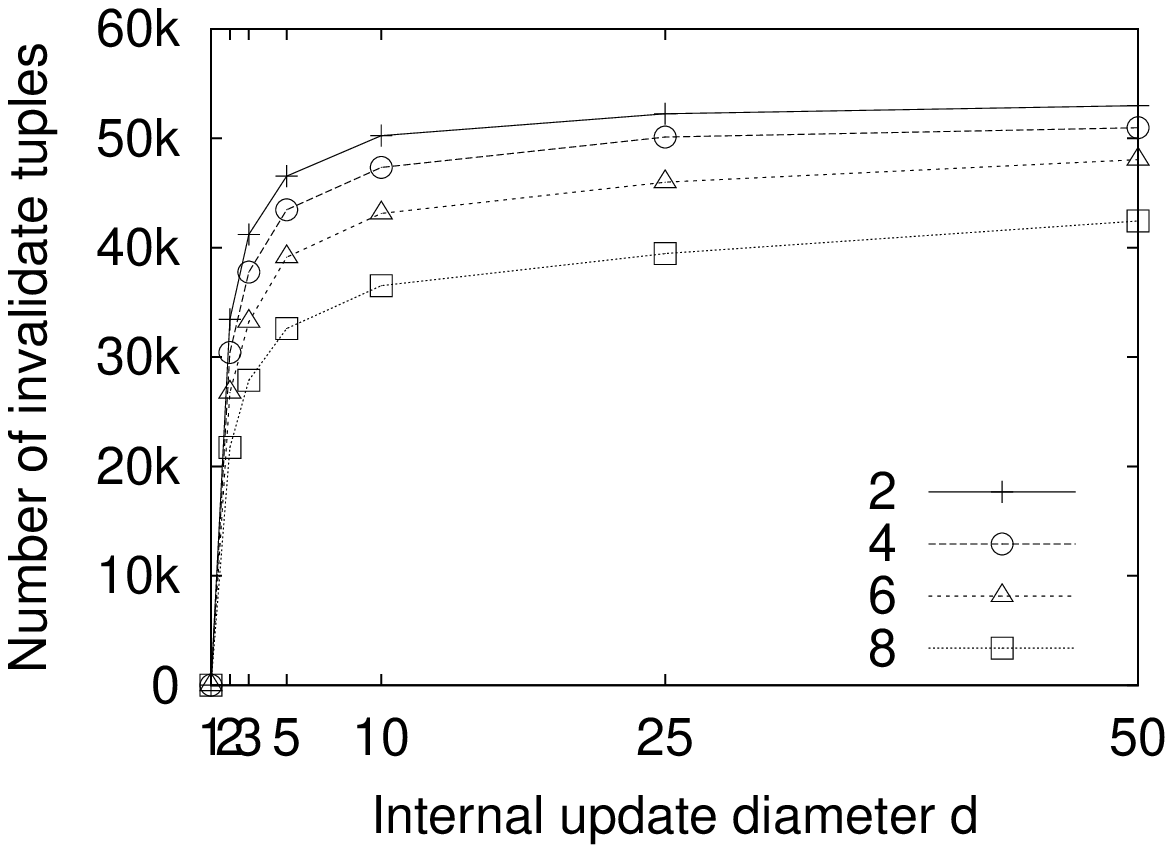}
\end{minipage}

} \vspace{-1em}\caption{Number of invalidate records}\vspace{-1em}
\end{figure}

In the next experiment, we show the invalidation of
\textit{m}-Invariance. We adopt the algorithm
in~\cite{xiao:m-invariance} to re-publish dataset and report the
total number of invalidate records.
As expected, the invalidate counts increase gradually as new
publication releases~(fig.~\ref{invalidate_totalTuples_time_Vs_m}).

Since the invalidation of \textit{m}-Invariance is cause by internal
updates, we test the invalidate record counts with respect to the
diameter~(fig.~\ref{invalidate_totalTuples_d_Vs_m}).The invalidate
counts increase dramatically with respect to the increase of
diameter.
Specially, when the diameter is 1, which means each value can not
update to be anyone else, there is no invalidate records. That
reconfirmed that the invalidation reason of \textit{m}-Invariance is
internal update.

\subsection{\textit{m}-Distinct Evaluation}

In this subsection, we will evaluate our solution from the following
aspects:

\subsubsection{Query Accuracy}
We test the query accuracy of anonymization data by answering
aggregate queries as follows:
\begin{tabbing}
\vspace{-3em}\small{SELECT COUNT(*)}\\

\small{FROM $T_i$}\\

\small{WHERE} \= \small{$Q_1>a_1$} \= \small{AND} \small{$Q_1<b_1$}\\

\>\> \small{...}\\

\> \small{$Q_4>a_4$} \> \small{AND} \small{$Q_4<b_4$}\\

\> \small{$S >a_5$} \> \small{AND} \small{$S <b_5$} \vspace{-5em}
\end{tabbing}
For each query, we configure its range as $|b_j-a_j|=\theta \cdot
|dom(A_j)|$, where $\theta \in(0,1]$. Apparently, a query with
larger $\theta$ will involve more records and return a larger
result.

The query error, which is the difference between the returned
results on $T_i$ and $T^{*}_i$, is $|R^{*}-R|/R^{*}$. $R$ is the
query result on $T_i$ and $R^{*}$ is the estimated result on
$T^{*}_i$. Specifically, the estimated result is the number of
possible records which are in QI-group $g$ and covered by the query.
Suppose the records in $g$ are uniformly distributed, the
probability of a record $t$ meets the query is the product of
probabilities that $t[A_j]$~($i=1,2,...,5$) is in the interval
$(a_j, b_j)$. Thus $R^{*}$ equals to the product of the records
count of $g$~(excluding counterfeit counts) and the above
probability.

\begin{figure}

\centering \subfigure[\small{Vs.
time~($\theta=50\%$)}\label{query_time_m_Vs_error}]{
\begin{minipage}[b]{0.22\textwidth}
\includegraphics[width=1\textwidth]{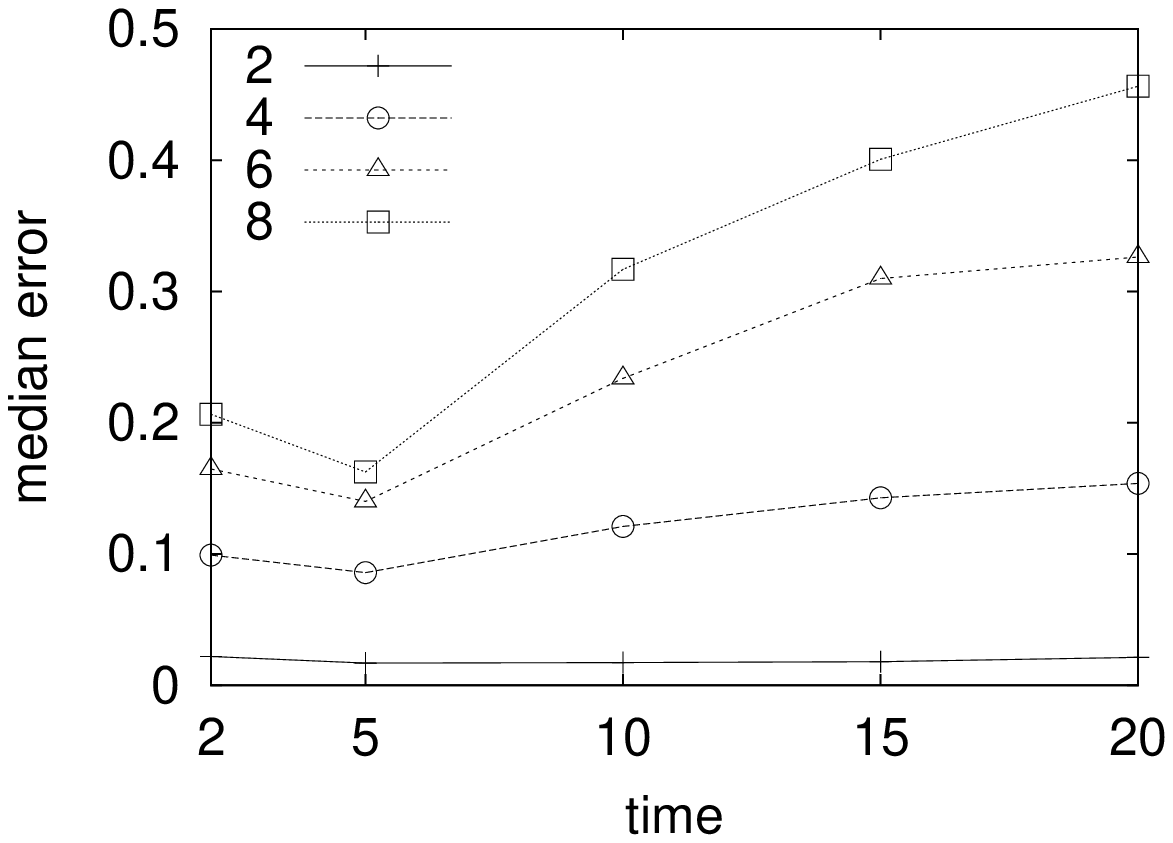}

\end{minipage}%
} \subfigure[\small{Vs.
$\theta$~($m=4$)}\label{query_range_Vs_error}]{
\begin{minipage}[b]{0.22\textwidth}

\includegraphics[width=1\textwidth]{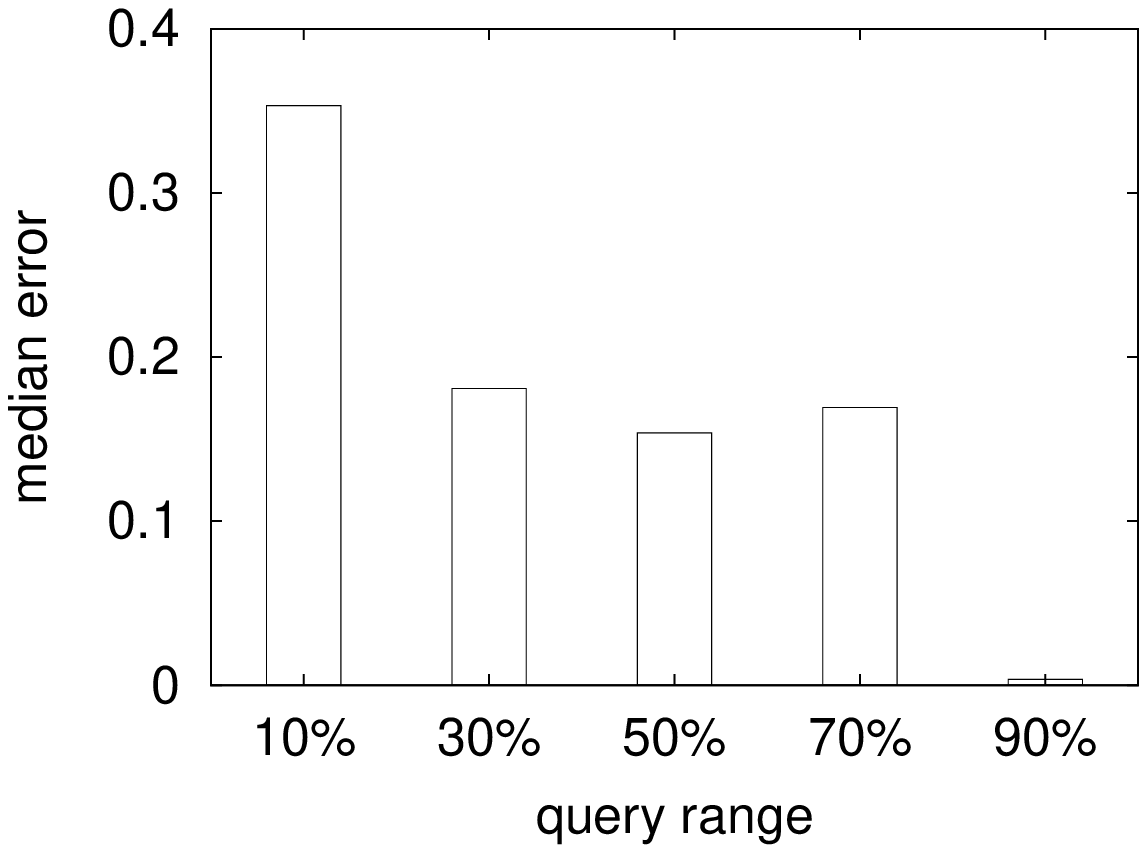}
\end{minipage}

} \vspace{-1em}\caption{Query error}\vspace{-1em}
\end{figure}

In this experiment, we randomly generate 10,000 queries and report
the median error. Fig.~\ref{query_time_m_Vs_error} shows the median
errors for different time and \textit{m}. The median error increases
smoothly as time evolves, because the new inserted records for a
QI-group are usually not as 'well' as the deleted ones; however, as
the result of \textit{2(4)}-Distinct demonstrated, the median error
will not increase anymore when re-publishing enough times.

\begin{figure}
\centering
\begin{minipage}[c]{0.22\textwidth}
\centering
\includegraphics[width=1\textwidth]{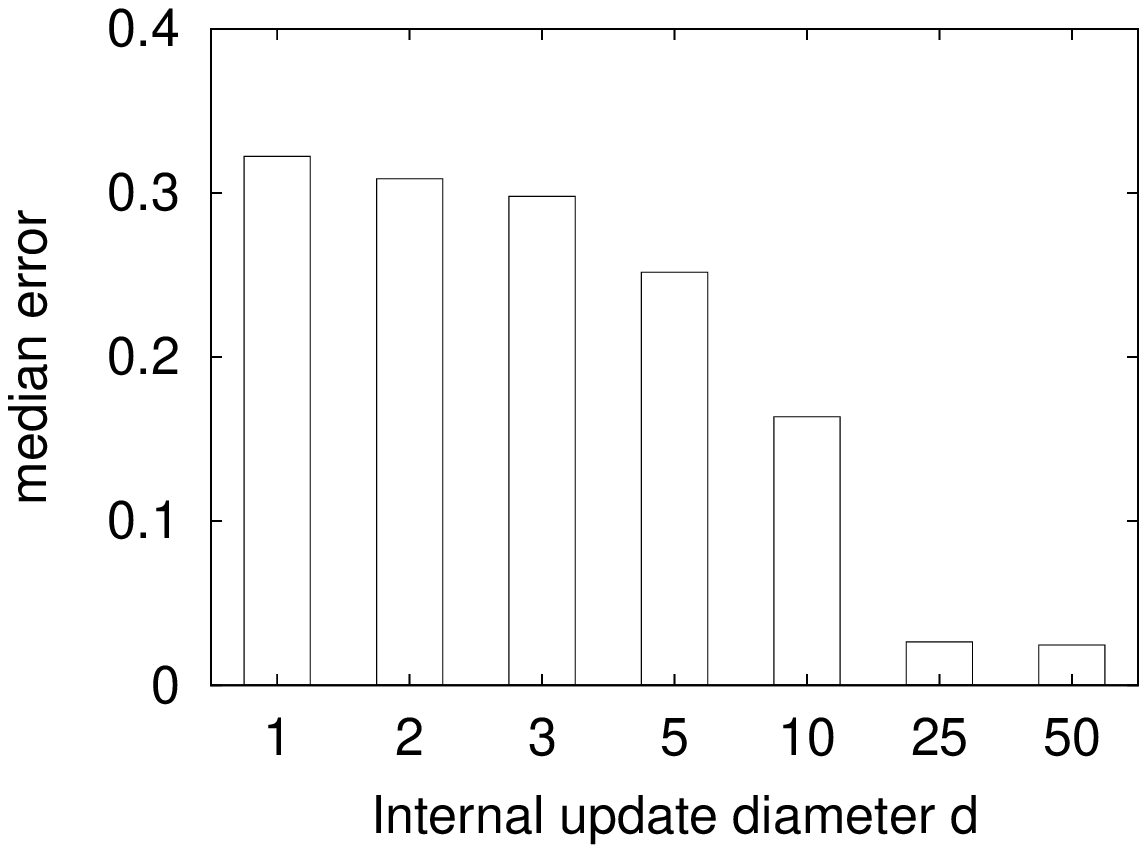}
\vspace{-2em}\caption{\small{Query error Vs. \textit{d}~($m=4,
\theta=50\%$)}\label{query_d_Vs_error}\vspace{-4em}}
\end{minipage}
\begin{minipage}[c]{0.22\textwidth}
\centering
\includegraphics[width=1\textwidth]{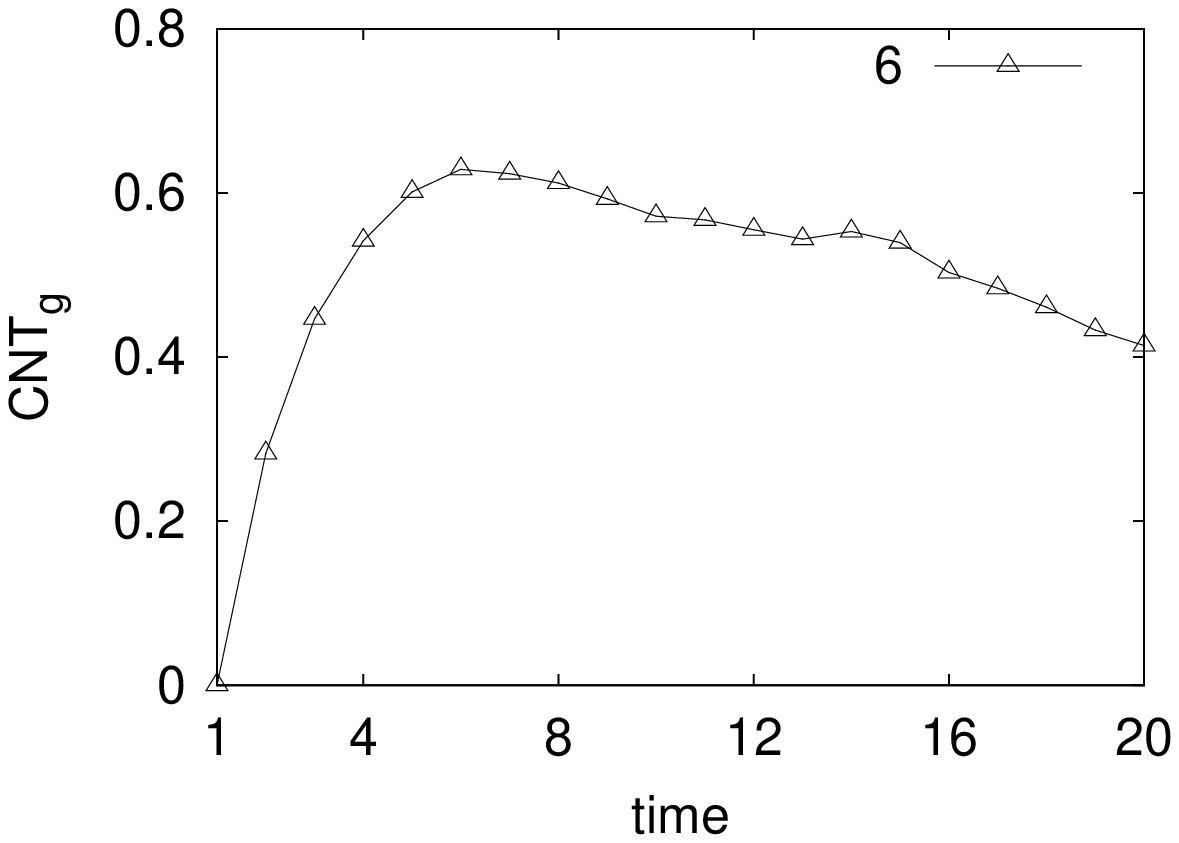}
\vspace{-2em}\caption{\small{ $CNT_g$ Vs.
time~($m=6,d=5$)}\label{c_count_Vs_time}\vspace{-4em}}
\end{minipage}%

\end{figure}

Fig.~\ref{query_range_Vs_error} shows the median error versus
different $\theta$ while $m=4$. As expected, the query with larger
range gets more accurate result, because when it covers more
records, the estimation are more close to the actual result. At
last, in fig.~\ref{query_d_Vs_error} we show the median error for
different internal update diameter. Since a larger diameter means
more flexible internal updates, which allows more records to be
assigned in a bucket and is more flexible to generate QI-groups, the
accuracy increases with $d$.

\subsubsection{Counterfeit Counts}
In this experiment we temporarily configure the delete amount of
each re-publication to be 3,000 because our initial setup leads to
zero counterfeit count in most of evaluations. We define the
measurement as the average counterfeit count per QI-group~(denote as
$CNT_g$).

Fig.~\ref{c_count_Vs_time} plots the $CNT_g$ versus time when $m=6$
and $d=5$. $CNT_g$ increases at the beginning because more existing
QI-groups have records been removed, that leads to a more urgency of
counterfeit records in the balance step. However, after the $7^{th}$
publication, $CNT_g$ decreases as the total number of counterfeit
records becomes more and more stabilized and the number of QI-groups
are always increasing.

\begin{figure}

\centering \subfigure[\small{Vs.
\textit{d}~($m=6$)}\label{c_count_Vs_d}]{
\begin{minipage}[b]{0.22\textwidth}
\includegraphics[width=1\textwidth]{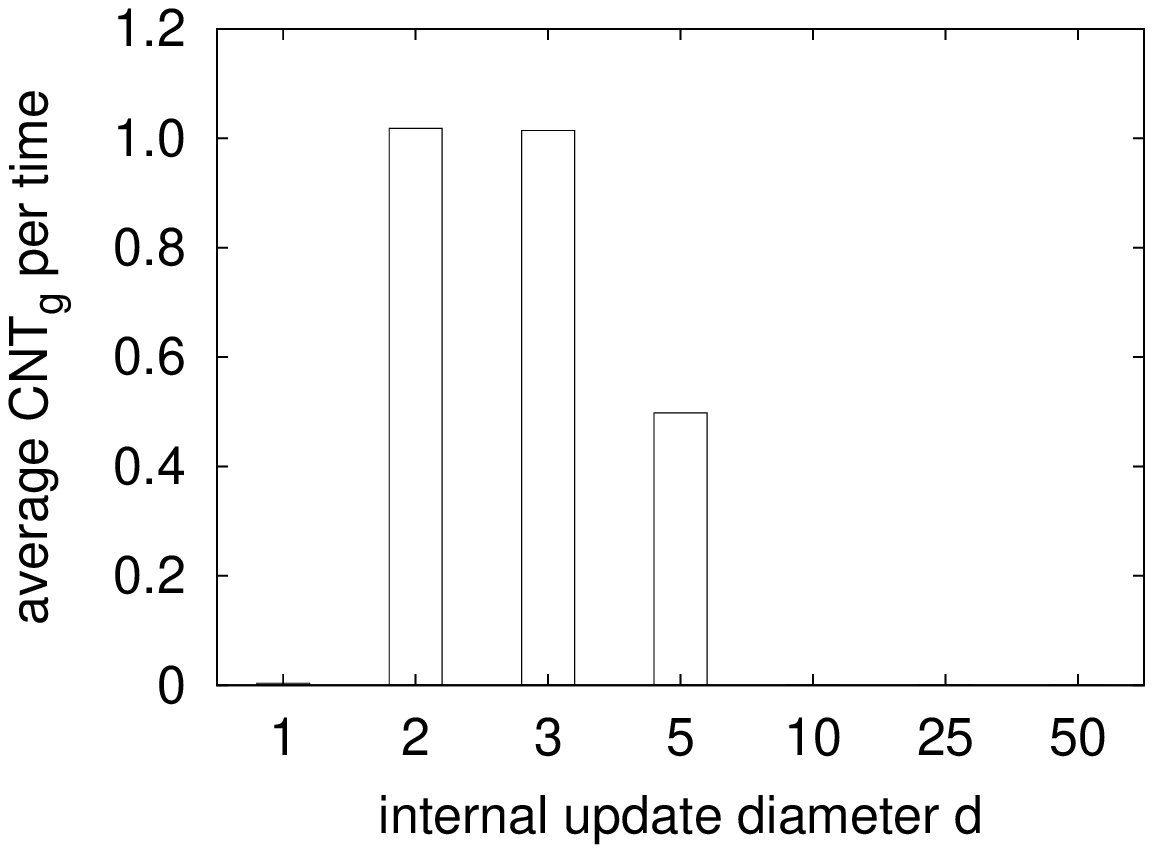}

\end{minipage}%
} \subfigure[\small{Vs. \textit{m}~($d=5$)}\label{c_count_Vs_L}]{
\begin{minipage}[b]{0.22\textwidth}

\includegraphics[width=1\textwidth]{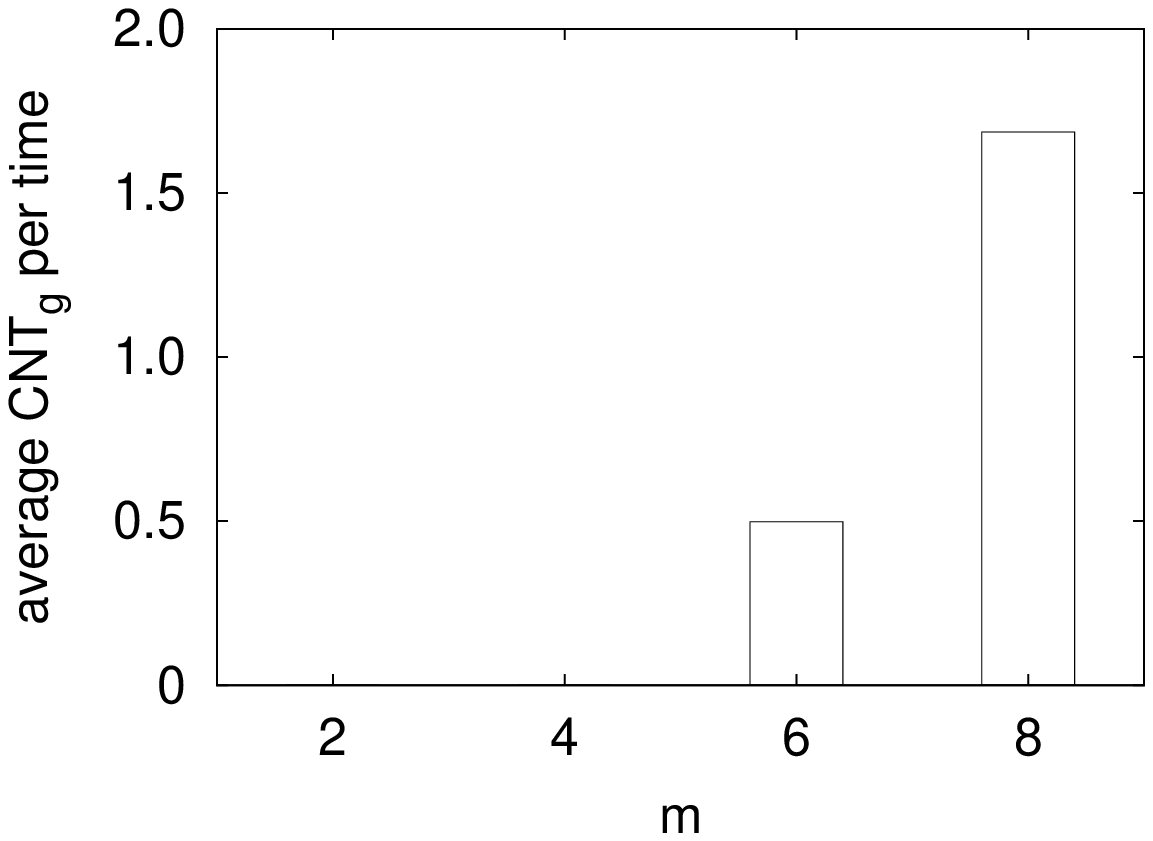}
\end{minipage}

} \vspace{-1em}\caption{average $CNT_g$ per time}\vspace{-1em}
\end{figure}

Then we fix $m=6$ and show the average $CNT_g$ per time with
different configuration of \textit{d}~(fig.~\ref{c_count_Vs_d}). It
is expected that a larger diameter makes the publication with less
counterfeit records in each group, because it also means more
flexible assignment for the inserted records.

In fig.~\ref{c_count_Vs_L} we set $d=5$ and measure the average
$CNT_g$ per time for different \textit{m}-Distinct. The result of
\textit{8}-Distinct is the largest but also smaller than 2.

\subsubsection{Computation Cost}
According to the experiment setup, the number of re-publication
records is incremental as time evolves. In order to accurately
measure the cost of a single re-publication procedure, we report the
average running time for re-publishing $T^{*}_2$.

Fig.~\ref{time_Vs_d} demonstrates the computation cost with
different internal update diameter. A larger diameter can lead to
higher data utility~(fig.~\ref{query_d_Vs_error}) as well as always
needs higher cost. Because in phase 2, there may be more optional
buckets for a new inserted records and more records will be assigned
in the same bucket, which will lead to more cost when splitting and
generating $QI$-groups in the last phase.

We also observe the cost comparing to different \textit{m}. The cost
decreases from 2 to 4 because a smaller \textit{m} has less number
of buckets and each bucket has more records, which need more split
operations to generate $QI$-groups. However, the overhead increases
from 4 to 8, that is because when \textit{m} becomes larger, the
major cost is how to find a legal split for a bucket with least
information loss, which is positively correlated to \textit{m}~(as
the analysis in section~\ref{phase_3}).

%% file: 7_related/related.tex
\section{Related Work}\label{section_7}
Most existing anonymization work is carried out on static datasets,
where records are inserted and/or deleted dynamically. Different
anonymization
principles~\cite{presence,anatomy,l-diversity,sweeney:k-anonymity,xiao:personalized,t-closeness}
have been proposed to preserve privacy and ensure the sensitive
information security from different perspectives. In addition to
resisting different kinds of disclosure
attacks~\cite{corruption,l-diversity,aggregate,minimality}, the
anonymization principles also struggle to achieve privacy
preservation with less information loss.
Many
algorithms~\cite{samarati:anonymity,mondrian,aggregate,high-dimensional,full-domain}
have are also been proposed to generalize datasets to meet the
principles with little overhead.

Relatively, the data re-publication has received less attention.
Wang and Fung~\cite{wang:sequential} first studied the problem of
securely releasing multi-shots of a static dataset. The main
challenge is the inference caused by joining between multiple
releases. They proposed a solution to properly anonymize the current
release
so as to control possible inferences.

The anonymization work on dynamic datasets was initiated in
~\cite{byun:incremental}. Byun et al. tackled the problem of
incremental dataset anonymization, where a dataset is updated by
only \textit{record insertion}.
Their solution
supports neither record deletion nor attribute value update.

Xiao and Tao~\cite{xiao:m-invariance} first conducted anonymization
on external dynamic datasets, which are updated by both
\textit{record insertion} and \textit{deletion}. The challenge lies
in that the inserted and deleted records may cause the disclosure
risk of both themselves and the remained records, and even lead to
the disclosure of individuals' sensitive values. Their solution,
called
\textit{m}-Invariance~\cite{xiao:m-invariance,dynamic-anonymization},
guarantees that each time the QI-group to which a record belongs
contains the same set of sensitive values.

In short, all existing work do not consider internal updates, and
their solutions are invalid for fully dynamic datasets. This
constitutes the task for our paper to solve.

\begin{figure}

\centering \subfigure[\small{Vs. \textit{d}
($m=4$)}\label{time_Vs_d}]{
\begin{minipage}[b]{0.22\textwidth}
\includegraphics[width=1\textwidth]{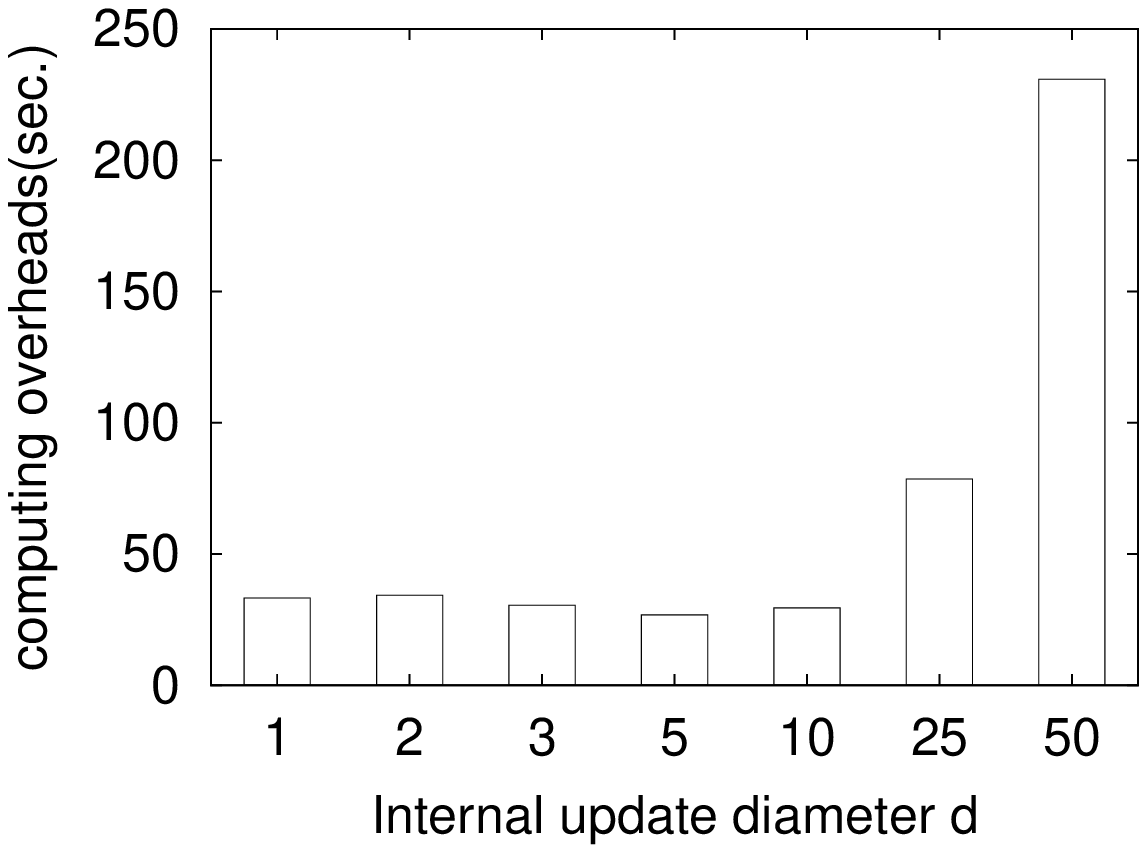}

\end{minipage}%
} \subfigure[\small{Vs. \textit{m}}\label{time_Vs_m}]{
\begin{minipage}[b]{0.22\textwidth}

\includegraphics[width=1\textwidth]{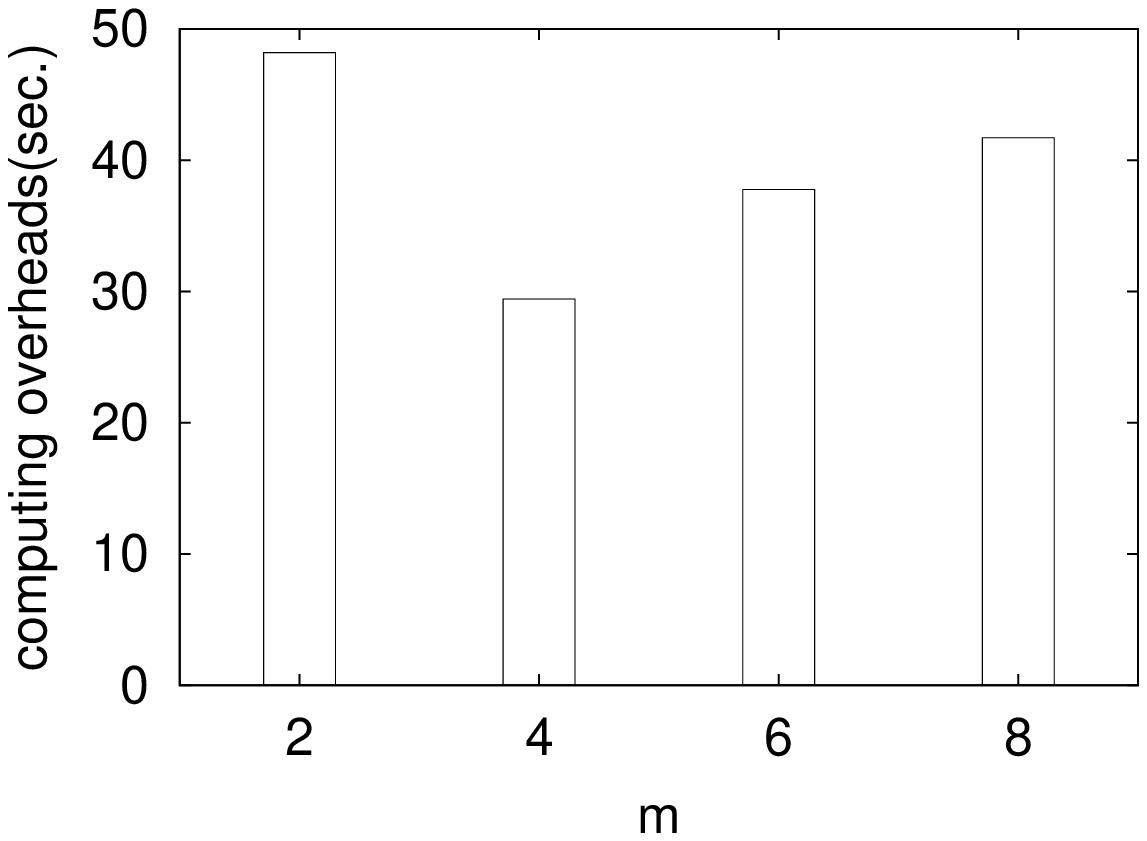}
\end{minipage}

} \vspace{-1em}\caption{Computation
cost}\label{Computation_cost}\vspace{-1em}
\end{figure}

%% file: 8_conclusion/conclusion.tex
\section{Conclusion}\label{section_8}
This paper challenges a new problem---enabling anonymization of
dynamic datasets with both internal updates and external updates.
For this goal, a novel privacy disclosure framework, which is
applicable to all dynamic scenarios, is proposed. A new
anonymization principle \textit{m}-Distinct and corresponding
algorithm are presented for anonymous re-publication of fully
dynamic datasets. Extensive experiments conducted on real world data
demonstrate the effectiveness of the proposed solution.

%% file: 9_appendix/appendix.tex
\section*{APPENDIX}

\textbf{Proof of lemma \ref{corollary}}. Suppose $t^{'}_{1},
t^{'}_{2}, ..., t^{'}_{I}$ are the sequential versions of record $t$
which exist in the corresponding releases of $T^{'}$. For any $1\leq
i\leq I$, we have $t^{'}_i\in T^{'}_i$. Regardless how to release
$T^{'}_{I+1}$, if we can prove $r_I(t^{'}_i)=r_{I+1}(t^{'}_i)$ holds
for any $1\leq i\leq I$, lemma \ref{corollary} must holds.

Derive from the condition presented in the lemma, we know that $t$'s
feasible sub-$SUG$ $G^{'}_{I+1}(V^{'},E^{'})$ can be constructed on
the basis of $G^{'}_I(V^{'},E^{'})$: add $|C_{I+1}|$ nodes which
represent the sensitive candidate set $C_{I+1}$ and draw edges from
every node in the last sensitive candidate set of
$G^{'}_I(V^{'},E^{'})$ to every new node.

Suppose $p^{'}_k$ is any feasible path in
$G^{'}_{I+1}(V^{'},E^{'})$, its weight can be represented as
$w(p^{'}_k)=w(p_k)w(v^{'}_{I,x_I},v^{'}_{I+1,x_{I+1}})w(v^{'}_{I+1,x_{I+1}})=\frac{1}{|C_{I+1}|}w(p_k)w(v^{'}_{I+1,x_{I+1}})$,
where $p_k$ is a feasible path in $G^{'}_I(V^{'},E^{'})$ and
contained by $p^{'}_k$; node $v^{'}_{I,x_I}$ and
$v^{'}_{I+1,x_{I+1}}$ are in $V^{'}_I$ and $V^{'}_{I+1}$
respectively and crossed by $p^{'}_k$.

According to equation \ref{eq3}, we have
$r_I(t^{'}_i)=\frac{\sum^{K_i}_{k^{'}=1}w(p_{k^{'}})}{\sum^{K}_{k=1}w(p_{k})}$.
Similarly, we have
$r_{I+1}(t^{'}_i)=\frac{\sum^{K_i\cdot|C_{I+1}|}_{k^{'}=1}w(p^{'}_{k^{'}})}{\sum^{K\cdot|C_{I+1}|}_{k=1}w(p^{'}_{k})}$.
According to the construction process stated above, in
$G^{'}_{I+1}(V^{'},E^{'})$, the number of total feasible
paths~($t^{'}_i$ related feasible paths) are $|C_{I+1}|$ times of
the count in $G^{'}_{I}(V^{'},E^{'})$. Collaborate
$r_{I+1}(t^{'}_i)$ with $p^{'}_k$, we have
$r_{I+1}(t^{'}_i)=\frac{\sum^{|C_{I+1}|}_{m^{'}=1}\sum^{K_i}_{k^{'}=1}w(p_{k^{'}})w(v^{'}_{I+1,x_{m^{'}}})}{\sum^{|C_{I+1}|}_{m=1}\sum^{K_i}_{k=1}w(p_{k})w(v^{'}_{I+1,x_{m}})}\\
=\frac{\sum^{K_i}_{k^{'}=1}\{w(p_{k^{'}})\sum^{|C_{I+1}|}_{m^{'}=1}w(v^{'}_{I+1,x_{m^{'}}})\}}{\sum^{K_i}_{k=1}\{w(p_{k})\sum^{|C_{I+1}|}_{m=1}w(v^{'}_{I+1,x_{m}})\}}$.

Since the weight sum of the nodes in $V^{'}_{I+1}$ is 1, we also
have $\sum^{|C_{I+1}|}_{m^{'}=1}w(v^{'}_{I+1,x_{m^{'}}})=1$ and
$\sum^{|C_{I+1}|}_{m=1}w(v^{'}_{I+1,x_{m}})=1$. Thus
$r_{I+1}(t^{'}_i)=\frac{\sum^{K_i}_{k^{'}=1}w(p_{k^{'}})}{\sum^{K}_{k=1}w(p_{k})}$
holds, which is also equal to $r_I(t^{'}_i)$. Hence Lemma
\ref{corollary} is proved.

 \textbf{Proof of lemma \ref{v=1}}. According
to equation \ref{eq3}, $r_n(t^{'}_i)=1$ implies that the weight sum
of feasible paths crossing the node represents $t^{'}[S]$ equals to
the total sum of all the feasible paths. Since every feasible path
must cross only one node in $V^{'}_i$ and each node must in at least
one feasible path, the above condition satisfied only when there is
only one node in $V^{'}_i$.

Similarly, $|V^{'}_i|=1$ directly implies $K_i=K$ and
$\sum^{K_i}_{k^{'}=1}w(p_{k^{'}})=\sum^{K}_{k=1}w(p_{k})$, which
will lead to $r_n(t^{'}_i)=1$. Hence lemma \ref{v=1} is proved.

\textbf{Proof of lemma \ref{v>=m}}. Suppose $t$ is any record
involved in the sequential release of $T$. If the releases are
\textit{m}-Distinct, there are at least $m$ distinct sensitive
values for each $QI$-group $t$ is in, because each release is
$m$-unique. That implies that in $t$'s $SUG$ $G_n(V, E)$, $|V_i|>m$
holds for any candidate node set.

Now if we can prove that the deduce from $G_n(V, E)$ to
$G^{'}_n(V^{'}, E^{'})$ will not delete any edge and node,
$|V^{'}_i|>m$ must hold for $G^{'}_n(V^{'}, E^{'})$ of $t$. Actually
this holds because the candidate sensitive set of $t_{i+1}$
$C(t_{i+1})$ must be a legal update instance of $C(t_{i})$, which
also means that any node in $V_i$ at least has an outgoing edge
connecting to a node in $V_{i+1}$ as well as any node in $V_{i+1}$
at least has an incoming edge from a node in $V_i$. Since this holds
for all $i$, no deletion will happen in the deduce procedure. Hence
the lemma is proved.

\textbf{Proof of lemma \ref{risk<=1/m}}. According to the definition
of \textit{m}-Distinct$^{*}$, if a sequential releases are
\textit{m}-Distinct$^{*}$, they must also be \textit{m}-Distinct and
lemma \ref{v>=m} hold.

Since $CUS_{\alpha} \cap CUS_{\beta}=\phi$ ($\alpha \neq \beta$)
holds for any two sensitive values in $t_{1}$'s candidate sensitive
set, in $t$'s $G^{'}_n(V^{'}, E^{'})$, there does not exist two
edges from $V^{'}_i$ to $V^{'}_{i+1}$ ($i=1$) cross the same node.
According to the transitivity of updates, the above property also
holds for $i=1,2,...,n$. Thus we can derive that there are at least
$m$ parallel feasible paths in $t$'s $G^{'}_n(V^{'}, E^{'})$.
Following the random world assumption, each feasible path has equal
weight and the disclosure risk for $t$'s any sensitive value is at
most $1/m$ (equation \ref{eq3}). Hence the re-publication risk is at
most $1/m$ and lemma \ref{risk<=1/m} is proved.